\documentclass[a4paper,12pt]{article}
\usepackage[intlimits]{amsmath}
\usepackage{hyperref}
\usepackage[russian,english]{babel}
\usepackage{euscript}
\usepackage{amssymb,amsmath,amsthm}
\usepackage{bbm}
\usepackage{esvect}
\usepackage{setspace}
\usepackage{epsfig}
\usepackage{caption}

\usepackage[T1]{fontenc}
\usepackage{fouriernc}
\usepackage{pstricks,pst-node}
\usepackage{color}
 
\newcommand{\bea}{\begin{equation}}
\newcommand{\eea}{\end{equation}}
\newcommand{\bear}{\begin{eqnarray}}
\newcommand{\eear}{\end{eqnarray}}
\newcommand{\bearr}{\begin{eqnarray*}}
\newcommand{\eearr}{\end{eqnarray*}}

\newcommand{\beal}{\begin{align}}
\newcommand{\eeal}{\end{align}}
\newcommand{\beall}{\begin{align*}}
\newcommand{\eeall}{\end{align*}}

\newcommand{\cf}{\mathcal{F}}
\newcommand{\tr}{\mathrm{tr}\,}
\newcommand{\CP}{\mathbf{C}\mathrm{P}}
\newcommand{\CC}{\mathbf{C}}
\newcommand{\dd}{\partial}
\newcommand{\im}{\mathrm{i}\,}
\newcommand{\dx}{\partial_x}

\newcommand{\as}{\mathrm{a}}
\newcommand{\bs}{\mathrm{b}}
\newcommand{\cs}{\mathrm{c}}
\newcommand{\xs}{\mathrm{x}}
\newcommand{\ys}{\mathrm{y}}
\newcommand{\zs}{\mathrm{z}}
\newcommand{\ol}{\frac{1}{L}}
\newcommand{\olsq}{\frac{1}{L^2}}
\DeclareMathAlphabet{\mathpzc}{OT1}{pzc}{m}{it}
\newcommand{\comment}[1]{}
\newcommand{\ug}{\mathfrak{u}}
\newcommand{\hg}{\mathfrak{h}}
\newcommand{\tgo}{\mathfrak{t}}
\newcommand{\sym}{\mathrm{Sym}}
\newcommand{\pfat}{\hat{\mathbb{P}}}

\begin{document}
\title {Haldane limits
via Lagrangian embeddings}
%and sigma models on flag manifolds}
\author {Dmitri Bykov\footnote{Emails:
dbykov@maths.tcd.ie, dbykov@mi.ras.ru}
\\
{\small{\it School of Mathematics,
Trinity College, Dublin 2, Ireland}} \\ {\small {\it Steklov
Mathematical Institute, Gubkina str. 8, 119991 Moscow, Russia \;}}}
\date{}
%\preprint{
   %       \smaller{\smaller{\smaller{TCDMATH 10-02}}}}
\maketitle 
\vspace{-0.75cm}
\begin{abstract}
In the present paper we revisit the so-called Haldane limit, i.e. a particular continuum limit, which leads from a spin chain to a sigma model. We use the coherent state formulation of the path integral to reduce the problem to a semiclassical one, which leads us to the observation that the Haldane limit is closely related to a Lagrangian embedding into the classical phase space of the spin chain. Using this property, we find a spin chain whose limit produces a relativistic sigma model with target space the manifold of complete flags $U(N)/U(1)^N$. We discuss possible other future applications of Lagrangian/isotropic embeddings in this context.
\end{abstract}

%\date {4.02.2011}

\section{Introduction}

Spin chains and sigma models are the two colossi of two-dimensional physics. Both have a long and rich history, although the sigma models are a much younger, and hence less understood, subject. By their very definition spin chains are finite-dimensional objects, and therefore many problems related to them are more easily formulated and in many cases may be solved numerically (although the simplicity of formulating a problem does not mean it is always easy to solve it analytically). On the other hand, sigma models are examples of interacting quantum field theories and share the common drawbacks of the latter --- namely, that generically they can only be formulated in the framework of perturbation theory, which leaves much to be desired: in each of the terms one encounters infinities, which have to be renormalized, and, as if it were not bad enough, it is not known whether the perturbation series can be made to converge by any reasonable means. The situation improved with the advent of integrable methods in two-dimensional quantum field theory. It turned out that these difficulties can be bypassed, if not resolved, if the theory at hand possesses an infinite number of commuting conserved ``charges''. Even more importantly, it was shown that the various continuum limits of the spin chains produce sigma model-like actions, and therefore the spin chains may serve as natural regularizers, which preserve the symmetries of the sigma models (for an important example of such approach see \cite{FR}). However, for a given spin chain there may exist different inequivalent continuous limits, and the models which arise as a result will differ substantially. For example, for the most common case of the $SU(2)$  spin chain with interactions of the form $\vec{S}_i \vec{S}_{i+1}$ the long-range fluctuations over the ferromagnetic vacuum are described by the so-called Heisenberg ferromagnet model (see Section \ref{ferro} below), which is a splendid model apart from being relativistic. An important result of \cite{H} was that the continuum limit around the antiferromagnetic configuration produces a \emph{relativistic} sigma model with target space the sphere $S^2$ (though with a ``theta-angle'' $\theta = \pi m$, $m$ being the integer characterizing the representation in which each site of the chain transforms).
%(even though with a $\theta$-term equal to $2 \pi S$)

The interest to the spin chains and sigma models especially increased after the advent of the AdS/CFT correspondence (for a review see \cite{rev1, rev2}). One of the examples of the AdS/CFT correspondence relates a supersymmetric conformal quantum field theory in a three-dimensional spacetime to a string sigma model with target space $AdS_4 \times \CP^3$. Investigating this sigma model and generalizing earlier results of \cite{AM}, we found \cite{Bykov} that a particular low-energy limit of this model produces a standard $\CP^3$ action, where the bosonic fields interact with a single Dirac fermion. The Lagrangian of that model may be written as follows:
\bea\label{final}
\mathcal{L} = \,\eta^{\alpha\beta}\, \overline{\mathcal{D}_\alpha z^j} \, \mathcal{D}_\beta z^j  \,+ \,i 
\overline{\psi} \gamma^\alpha\widehat{\mathcal{D}}_\alpha \psi +{g\over 4} (\overline{\psi} \gamma^{\alpha} \psi)^2,
\eea
where index \(j\) runs from 1 to 4, $\mathcal{D}_\alpha= \partial_\alpha-i\,\mathcal{A}_\alpha$, $\widehat{\mathcal{D}}_\alpha= \partial_\alpha+2\, i\,\mathcal{A}_\alpha$. $\mathcal{A}_\alpha$ is a $U(1)$ gauge field without a kinetic term --- it can be integrated out to provide the conventional Fubini-Study form of the action. Besides, in (\ref{final}) the $z^j$ fields are restricted to lie on the $S^7 \subset \mathbb{C}^4$:
\bea\label{final2}
\sum\limits_{j=1}^4 |z^{j}|^2=R^2
\eea
The model defined by (\ref{final}), (\ref{final2}) has two coupling constants, the ``radius'' $R$ and the four-fermion coupling $g$. It is not known whether this model is integrable, but the methods developed in this paper may, with a bit of luck, lead to a resolution of this question in the future.

In the paper we will argue that the generalized Haldane limits of the models described by Hamiltonians (\ref{aff}) and (\ref{xxx}) (see below) are relativistic sigma models with target spaces $\CP^N$ and $\mathcal{F}_N$ (flag manifold) respectively. The $\CP^N$ sigma model was obtained in this manner for the first time by Affleck \cite{Affleck} and, in a three-dimensional setting by Read and Sachdev \cite{RS}. To our knowledge, the $SU(N)$ flag sigma model has not been obtained yet in this manner (see, however, \cite{indian}, whose authors obtained a real flag manifold $O(3)/Z_2^3$, and \cite{salam}\footnote{I would like to thank K.Zarembo for pointing out this reference to me.} for related constructions). In the present paper we will also develop a general framework for the ``Haldane limits'' and explain with the help of two examples (Observations 1 and 2 below) that they are closely related to the isotropic embeddings of certain manifolds.

The paper is organized as follows. In Section \ref{chains} we introduce the Hamiltonians of the spin chains that we will be analyzing. In Section \ref{coh} we recall the coherent state formalism and in Section \ref{quantsph} we apply it, as a pedagogical exercise and an introduction to what follows, to a quantum mechanical model with phase space the sphere $S^2$. Section \ref{pathint} is dedicated to a generalization of this discussion to the spin chain setup. In particular, in \ref{xxxchain} we build a path integral for the $XXX$ spin chain and in \ref{ferro} we analyze the continuum limit around the ferromagnetic vacuum. Sections \ref{anti} and \ref{cl} are at the heart of the paper --- there we analyze the expansion around the \emph{anti}ferromagnetic configuration. First, in Section \ref{aff1}, we rephrase in our language the results of Affleck related to the spin chain described by the Hamiltonian (\ref{aff}). Then in Sections \ref{antifer} and \ref{cl} we generalize this result to the case of the antiferromagnetic configuration of the Hamiltonian (\ref{xxx}). In particular, in Section \ref{antifer} we find out what the antiferromagnetic vacuum of the spin chain (\ref{xxx}) looks like classically. Section \ref{expansion} is devoted to the expansion of the action around this vacuum configuration. As a result we obtain the $SU(3)$ flag sigma model action (\ref{flagaction}), which is one of the main results of the present paper. In Section \ref{flagmetr} we elaborate on what the most general $SU(N)$-invariant metric on a flag manifold looks like. In Section \ref{suN} we discuss the generalization of our $SU(3)$ result to the case of $SU(N)$, and we discuss possible applications of the flag sigma model in condensed matter physics, in particular with regards to the so-called ``trimerization''. The discussion in Section \ref{disc} is dedicated to an overview of possible ways and directions of extending this line of research in the future. The paper contains three appendices. In Appendix \ref{app1} we give the basic definitions if the permutation and trace operators. Appendix \ref{appLL} is dedicated to a general discussion of the Landau-Lifshitz models, some examples of which are encountered in the main text of the paper. Last but not least, in Appendix \ref{appflag} we prove that the flag manifold $\cf_N$ may be isometrically embedded into the product $(\CP^{N-1})^{\times N}$ as a Lagrangian submanifold.

\section{The \(SU(N+1)\) spin chains}\label{chains}

In the following we will frequently encounter the complex projective space $\CP^N$ viewed as a homogeneous space of $SU(N+1)$. For this reason we prefer to write $SU(N+1)$ instead of the arguably more easily readable symbol $SU(N)$. In this paper we consider two families of spin chains having $SU(N+1)$ global symmetry. Members of each family are parametrized by the representation in which each site of the spin chain transforms. Moreover, we have no intention to elaborate on the most general situation possible, but rather wish to present to the reader a couple of clear and representative examples. For this reason the representations considered in this paper will be symmetric powers of the fundamental (and/or anti-fundamental) representation $\sym(V^{\otimes m})$. The simplest representatives of the two families are defined by the following Hamiltonians (in the following \(L\) is the length of the spin chain):
\bea\label{aff}
H_1 = \sum\limits_{i=1}^{L} \, \mathrm{Tr}_{i,i+1},
\eea
where $\mathrm{Tr}_{i,i+1}$ is the trace operator, and
\bea\label{xxx}
H_2=\sum\limits_{i=1}^{L} \, (P_{i,i+1}+P_{i,i+2}),
\eea
where \(P_{i,i+1}\) is the permutation operator. Both $\mathrm{Tr}_{i,i+1}$ and $P_{i,i+1}$ act on the product of two \(N+1\)-dimensional vector spaces \(\mathbf{C}^{N+1} \otimes \mathbf{C}^{N+1}\). Nevertheless, there is an important difference between Hamiltonians $H_1$ and $H_2$, which will play a role in the foregoing discussion. It comes from the requirement that the spin chain should possess $SU(N+1)$ symmetry. For this to be the case, the action of the group on the vector spaces --- sites of the spin chains --- is different for (\ref{aff}) and (\ref{xxx}). In the spin chain (\ref{aff}) the consecutive sites transform in contragradient representations (say, site $i$ in the fundamental $V$, and sites $i\pm 1$ in the antifundamental $V^\ast$). In the spin chain (\ref{xxx}) the representations at each site should be the same, and we take it to be the fundamental $V$.

It is well-known (and easy to check) that the operators $\mathrm{Tr}$ and $P$ entering the above Hamiltonians have simple matrix representations. Recall that in the Lie algebra $u(N+1)$ there's an invariant Killing scalar product, which is usually denoted by $\kappa$. Let $\lambda_n$ be a basis of generators in the fundamental representation, and $\bar{\lambda}_n$ --- the conjugate matrices, generating the antifundamental representation. Then
\bear
&&\sum\limits_{n,m} \kappa_{nm}\; \lambda_n \otimes \lambda_m=P \qquad \textrm{and}\\
&&\sum\limits_{n,m} \kappa_{nm}\; \lambda_n \otimes \bar{\lambda}_m=\mathrm{Tr}\;.
\eear
The generalization of the Hamiltonians (\ref{aff}) and (\ref{xxx}) to the case where the sites are in symmetric powers $\sym V^{\otimes m}$ of the fundamental representation is as follows. Take the generators $\Lambda^{(i)}_n(m)$ in these particular representations at each site $i$ (clearly, $\Lambda^{(i)}_n(1)=\lambda_n^{(i)}$) and build the following analog of the permutation operator: $ \pfat^{(i)}={1\over m}\sum\limits_n \Lambda^{(i)}_n(m) \bigotimes \Lambda^{(i+1)}_n(m)$. The sum of this density over all sites $i$ gives the generalization of the permutation operator relevant to our discussion. It is not difficult to convince oneself, however, that this is \emph{not} the same as the permutation operator on $\sym (V^{\otimes m}) \otimes \sym (V^{\otimes m})$. Indeed, the generators $\Lambda_n(m)$ (we omit the index $i$ for the moment) are the restrictions to the relevant subspace (necessary due to the symmetrization) of the following:
\bea
\Lambda_n(m)=\sum\limits_{s=1}^m 1\otimes...\otimes \underset{\underset{\textrm{s-th place}}{\uparrow}}{\lambda_n}\otimes...\otimes 1,
\eea
where there are $m$ factors and $\lambda_n$ stands in position $s$. This means that $\pfat^{(i)}$ looks as follows:
\bea\label{pfat}
\pfat^{(i)}={1\over m}\sum\limits_n \Lambda^{(i)}_n(m) \bigotimes \Lambda^{(i+1)}_n(m)=
{1\over m}\sum\limits_{s,t=1}^m \sum\limits_{n} (1\otimes...\otimes \underset{\underset{\textrm{s-th place}}{\uparrow}}{\lambda_n}\otimes...\otimes 1) \bigotimes (1\otimes...\otimes \underset{\underset{\textrm{t-th place}}{\uparrow}}{\lambda_n} \otimes...\otimes 1).
\eea
We know what the sum over $n$ in the above expression is (for fixed $s$ and $t$) --- it is the permutation operator between site $s$ of the first factor in $\sym (V^{\otimes m}) \otimes \sym (V^{\otimes m})$ and site $t$ in the second factor. Thus,
\bea
\pfat^{(i)}={1\over m}\sum\limits_{s,t=1}^m P_{s(i),\,t(i+1)}
\eea
The summation is necessary essentially to ensure that after permuting individual sites of a symmetrized product it remains symmetric.

\emph{Example}. Let us give a brief example. Suppose $m=2$. Then $\pfat^{(i)}$ acts as follows on a tensor $T_{ij|kl}$, symmetric in the first two and last two indices respectively:
\bea
\pfat(T_{ij|kl})={1\over 2}(T_{kj|il}+T_{ik|jl}+T_{lj|ki}+T_{il|kj})
\eea
If we prefer to use polynomials instead of symmetric tensors, we may write
\bear
\pfat(z_i z_j w_k w_l)={1\over 2}(z_k z_j w_i w_l+z_i z_k w_j w_l+z_l z_j w_i w_k+z_i z_l w_j w_k)=\\ \nonumber
={1\over 2} (z_k w_l+z_l w_k)  ( z_j w_i +z_i w_j),
\eear
the second expression vividly demonstrating the symmetry $i \leftrightarrow j$ and $k \leftrightarrow l$, which therefore proves that the action of the operator is well-defined.

The whole point of this lengthy and pedantic discussion is to convince the reader as to how $\pfat^{(i)}$ acts on the coherent states defined below:
\bea
\pfat [(a\circ \bar{z})^m (b\circ \bar{w})^m]=m \,(a\circ \bar{z})^{m-1} \,(b\circ \bar{z}) \;\;(b\circ \bar{w})^{m-1} \,(a\circ \bar{w})
\eea

\emph{Remark}. For the case of $SU(2)$ the fundamental and anti-fundamental representations are equivalent. This means that there exists a matrix $C$ such that
\bea
\bar{\lambda}_n=C \lambda_n C^{-1},\quad n= 1, 2, 3.
\eea
Let us construct a matrix $\hat{C}=1\otimes C \otimes 1 \otimes ...$ with a total of $L$ factors, $1$'s standing in the sites of the fundamental representation and $C$'s in the sites of the anti-fundamental one. Then we see that $P$ and $\mathrm{Tr}$ are conjugate to each other:
\bea
\mathrm{Tr}=\hat{C} P \hat{C}^{-1} .
\eea

\section{Coherent states}\label{coh}

It is a fact from algebraic geometry that any holomorphic line bundle over \(\CP^N\) is a (perhaps inverse) power of the tautological bundle\footnote{Sometimes this fact is formulated as the statement that the Picard group has a single generator: \(\mathrm{Pic}(\CP^N) \simeq \mathbf{Z}\).}. The line bundles over \(\CP^N\) are thus denoted by \(\mathcal{O}(m)\), where \(m\in\mathbf{Z}\) is the degree, or Chern number, of the bundle.

The coherent states are by definition particular sections of these line bundles \(\mathcal{O}(m)\). For a given \(m\) there's a very explicit realization of the vector space of such sections \(\Gamma(\mathcal{O}(m))\) --- these are the homogeneous polynomials of degree \(m\) of the \(N+1\) homogeneous coordinates of \(\CP^N\). There's a natural scalar product on \(\Gamma(\mathcal{O}(m))\). It is given by the following formula:
\bea\label{scalprod}
(f,g)=\,\int\; \overline{f(z)} \; g(z) \; \frac{d\mu(z,\bar{z})}{(\sum\limits_{i=1}^{N+1}\,|z_i|^2)^m},
\eea
where \(d\mu\) is the volume form on \(\CP^N\), which in homogeneous coordinates can be described as follows. Denote by $\tilde{\omega}$ the following holomorphic $N$-form:
\bea
\tilde{\omega} = \epsilon_{i_1,i_2, ..., i_{N+1}} z_{i_1}\; dz_{i_2} \wedge ...\wedge dz_{i_{N+1}}
\eea
and by $\overline{\tilde{\omega}}$ the conjugate one. Then the volume form looks as follows (notice that it is $U(N+1)$-invariant by construction):
\bea\label{cpvol}
(d\mu)_{\CP^N} = \omega_{FS}^N = \frac{\tilde{\omega} \wedge \overline{\tilde{\omega}}}{(\sum\limits_{k=1}^{N+1}\,z_k \bar{z}_k)^{N+1}}
\eea

%%%%%%%%%%%%%%%%%%%%%%%%%%%%%%%%%%%%%%%%%%%%%%%%%%%%%%%%%%%%%%%%%%%%%%%%%%%%%%%%%%%%%%%%%
\comment{
Denote by $\Omega_i$ the following holomorphic $N$-form:
\bea
\Omega_i = dz_1\wedge dz_2 ... \wedge \widehat{dz_i}\wedge...\wedge dz_{N+1},
\eea
where the hat means that the corresponding factor should be omitted. $\bar{\Omega}_i$ means a similar form, but with the $z$'s replaced with their conjugates. Then the volume form looks as follows:
\bear\nonumber
(d\mu)_{\CP^N}&=&\omega_{FS}^N=\sum\limits_{i,j} \Omega_i \wedge \bar{\Omega}_j \times \{\textrm{det of }\;ij\textrm{-th minor of the Hermitian Furbini-Study matrix}\; \omega_{FS} \} =\\ \label{cpvol}
&=&\sum\limits_{i,j} (-1)^{i+j} \frac{z_i \bar{z}_j}{(z_k \bar{z}_k)^{N+1}} \Omega_i \wedge \bar{\Omega}_j
\eear
}
%%%%%%%%%%%%%%%%%%%%%%%%%%%%%%%%%%%%%%%%%%%%%%%%%%%%%%%%%%%%%%%%%%%%%%%%%%%%%%%%%%%%%%%%%

It is a fact, characteristic of the projective space, that if one rescales all the $z$'s by a \emph{function}, the Fubini-Study form $\omega_{FS}$ as well as the volume form $d\mu$ will not change. An important consequence of this is that the integrand in (\ref{scalprod}) does not change under such rescalings. This means in particular, that one may choose a convenient ``gauge'', for example we can set $z_{N+1}=1=\bar{z}_{N+1}$ (in what follows we will mainly use this ``gauge''). In this case the volume form (\ref{cpvol}) simplifies:
\bea
(d\mu)_{\CP^N}=\frac{dz_1\wedge...\wedge dz_N \wedge d\bar{z}_1\wedge...\wedge d\bar{z}_N}{(1+\sum\limits_{k=1}^N \,z_k \bar{z}_k)^{N+1}}
\eea

For a given vector $|w\rangle$ of the representation $\hat{V}$ of the group $G$ coherent states are by definition \cite{Perelomov} an overcomplete basis for this representation, which is formed by the vectors in the orbit of the group $G$, containing $|w\rangle$. If $V$ is the fundamental representation of $SU(N+1)$, then the space $\sym(V^{\otimes m})$ may be identified with the space of all polynomials of degree $m$ on $N+1$ variables $z_i,\;i=1\,...\,N+1$. In order for the orbit to be as simple as possible we will take $|w\rangle= z_1^m$ --- one of the highest weights --- as element of the orbit\footnote{For a very clear exposition of these properties see \cite{FH} and \cite{Saemann}.}.
\comment{ In this case the weight of this state, or in other words the highest weight of the representation $\sym(\hat{V}^{\otimes n})$, is orthogonal to some simple roots (it lies on the border of some Weyl chamber clarify!!!???)}
 In this case the orbit is $\CP^N$ --- the smallest available one. Any element of the orbit may be written as follows\footnote{This notation is borrowed from \cite{Berezin}, who used coherent states of this sort to describe the quantization of a sphere $S^2\sim \CP^1$ --- the simplest homogeneous K\"{a}hler (symplectic) manifold.}:
\bea
\phi_{\bar{v}}(z)=(\sum\limits_i \, z_i \bar{v}_i)^m
\eea
The fact that the system is (over)complete means that the following fundamental identity is valid:
\bea\label{unity}
\textrm{(partition of unity)}\quad\quad \int d\mu(v,\bar{v}) \, \frac{\phi_{\bar{v}}(z)\, \phi_{v}(\bar{z})}{(\phi_{\bar{v}},\phi_{\bar{v}})}=1
\eea

%\subsection{Relation to Schwinger-Wigner quantization}

\section{The quantum sphere $S^2$}\label{quantsph}

Let us consider in detail the case of $su(2)$ \footnote{General results along a similar line of reasoning were obtained in \cite{AFS}}. When written using a single inhomogeneous coordinate, the coherent states are:
\bea
\phi_{\bar{v}}(z)=(1+z \bar{v})^m
\eea
The kernel of a generic operator $\hat{A}$ may be obtained using a standard construction
\bea
A(z,\bar{v})=\frac{(\hat{A}\phi_{\bar{v}},\phi_{\bar{z}})}{(\phi_{\bar{v}},\phi_{\bar{z}})}
\eea
Let us now regard the basis vectors $1, z$ as eigenvectors of the operator $\sigma_3$ with eigenvalues $1, -1$ respectively. Then the kernel of $\hat{A}=\hat{T}_3=
\sum\limits_{s=1}^m 1\otimes ... \otimes \underset{\underset{\textrm{s-th place}}{\uparrow}}{\sigma_3} \otimes ... \otimes 1$ is
\bea\label{gen}
T_3(q,\bar{v})=\frac{\left(m \,(1-z \bar{v})(1+z \bar{v})^{m-1},(1+z \bar{q})^m\right)}{(1+z \bar{v},1+z \bar{q})}=m\, \frac{1-q\bar{v}}{1+q \bar{v}}
\eea
We will now present the derivation of the matrix elements (between coherent states) of the ``evolution operator'' $\hat{U}=e^{-i\alpha \hat{T}_3}$. Of course, in this finite-dimensional case this is merely a pedagogical exercise, since clearly the action of $\hat{U}$ on a coherent state simply gives
\bea
\hat{U}\,\phi_{\bar{v}}=(e^{-\im\alpha}+e^{\im\alpha} z \bar{v})^m ,
\eea
so the corresponding matrix element is easily calculated:
\bea\label{evopans}
U(q,\bar{v})=\left(\frac{e^{-\im\alpha}+e^{\im\alpha} q \bar{v}}{1+q \bar{v}}\right)^m
\eea
As is standard in path integral calculations \cite{SF}, in order to write a path integral representation for a matrix element $U(q,\bar{v})$, we need to know the matrix elements of the generator (\ref{gen}), and then we need to split the ``time'' interval $\alpha$ into $K$ subintervals of length $\frac{\alpha}{K}$ and use the formula
\bea\label{U}
\hat{U}=\underset{K\to\infty}{\textrm{lim}} (1-\frac{\im\alpha}{K} \hat{T}_3)^K
\eea
Let us write down a path integral for (\ref{U}), using (\ref{unity}) and denoting $\hat{\tau} = 1+\frac{i\alpha}{K} \hat{T}_3$:
\bear\label{Upath1}
U(q,\bar{y})&=&\frac{(\hat{U}\phi_{\bar{y}},\phi_{\bar{q}})}{(\phi_{\bar{y}},\phi_{\bar{q}})}=\\ \nonumber
&=& \underset{K\to\infty}{\textrm{lim}} \int\;\prod\limits_{i=1}^{K-1} d\mu(z_i, \bar{z}_i)\;\; \tau(q,\bar{z}_{K-1}) \tau(z_{K-1},\bar{z}_{K-2})...\tau(z_2,\bar{z}_1) \tau(z_1,\bar{y}) \times \\ \nonumber &&\times \;
\frac{    (\phi_{\bar{y}},\phi_{\bar{z}_1})     (\phi_{\bar{z}_1},\phi_{\bar{z}_2})     \,...\,       (\phi_{\bar{z}_{K-2}},\phi_{\bar{z}_{K-1}})      (\phi_{\bar{z}_{K-1}},\phi_{\bar{q}})}{(\phi_{\bar{y}},\phi_{\bar{q}})     (\phi_{\bar{z}_1},\phi_{\bar{z}_1})     \,...\,       (\phi_{\bar{z}_{K-1}},\phi_{\bar{z}_{K-1}})}
\eear
To complete the derivation we use the following formulas:
\bearr
\tau(z_{k+1},\bar{z}_k)=1-m\, \frac{\im\alpha}{K}\; \frac{1-z_k \bar{z}_{k+1}}{1+z_k \bar{z}_{k+1}},\\
(\phi_{\bar{z}_{k}},\phi_{\bar{z}_{k+1}})=(1+z_k \bar{z}_{k+1})^m
\eearr
Then (\ref{Upath1}) takes the form
\bea\label{Upath2}
U(q,\bar{y})=  \underset{K\to\infty}{\textrm{lim}} \int\;\prod\limits_{i=1}^{K-1} d\mu(z_i, \bar{z}_i)\;\left(\frac{1+z_{K-1} \bar{y}}{1+q \bar{y}}\right)^m\;\prod\limits_{j=0}^{K-2} \left(\frac{1+z_j \bar{z}_{j+1}}{1+z_{j+1} \bar{z}_{j+1}}\right)^m \;\;\prod\limits_{j=0}^{K-1}\;\left( 1-m\,\frac{i\alpha}{K}\; \frac{1-z_j \bar{z}_{j+1}}{1+z_j \bar{z}_{j+1}} \right),
\eea
where $z_{0}=q$ and $\bar{z}_K=\bar{y}$. We now want to ``take the limit'' in the formula (\ref{Upath2}), assuming that $z_{i+1}-z_i \sim \frac{1}{K}$ (for a justification of this procedure see \cite{ZJ}). In order to do it we need to write the factors $\frac{1+z_j \bar{z}_{j+1}}{1+z_{j+1} \bar{z}_{j+1}}$ in the following form:
\bea
\frac{1+z_j \bar{z}_{j+1}}{1+z_{j+1} \bar{z}_{j+1}}=\left(1-\frac{(z_{j+1}-z_j) \bar{z}_{j+1}}{1+z_{j+1} \bar{z}_{j+1}}\right)\simeq 1-{1\over K}\frac{\dot{\bar{z}}_{j+1} \,z_{j+1}}{1+z_{j+1} \bar{z}_{j+1}},\quad j=0,1, ..., K-2 .
\eea
Then we obtain
\bea\label{evop}
U(q,\bar{y})=\int\limits_{\substack{ z(0)=q\\ \bar{z}(1)=\bar{y}}}\; \prod\limits_{t\in [0,1]} d\mu(z(t),\bar{z}(t))\;\left(\frac{1+z(1)\, \bar{y}}{1+q \bar{y}}\right)^m\;\exp{\left(-m\,\int\limits_0^1 \,dt\,\frac{\dot{\bar{z}} \,\bar{z}}{1+z\bar{z}}-m\,\im \,\alpha \int\limits_0^1 dt\, \frac{1-z\bar{z}}{1+z\bar{z}}\right)}
\eea
Let us elaborate on what the two terms in the exponent
\bea\label{lagr}
\mathcal{L}= \im \frac{\dot{z} \,\bar{z}}{1+z\bar{z}}-\alpha \frac{1-z\bar{z}}{1+z\bar{z}}
\eea
in (\ref{evop}) are. The first one $j=\frac{dz \,\bar{z}}{1+z\bar{z}}$ is the ``K\"{a}hler current'', in other words the connection in a fibre bundle over $\CP^1$, whose derivative produces the K\"{a}hler form: $dj=\frac{dz\wedge d\bar{z}}{(1+z\bar{z})^2}=\omega_{FS}$. The second term $H=\frac{1-z\bar{z}}{1+z\bar{z}}$ is the Hamiltonian. $z$ and $\bar{z}$ are the stereographic coordinates on the sphere $\CP^1$, and the equations of motion following from the Lagrangian (\ref{lagr}) describe the rotation of the sphere around its $z$-axis (the one orthogonal to the plane of the stereographic projection). Let us write out the e.o.m. which follow from the Lagrangian (\ref{lagr}):
\bea \label{harmosc}
\im \dot{z}=-2\alpha z ,\quad 
\im \dot{\bar{z}}=2\alpha \bar{z} .
\eea
We see that these are nothing but the equations of harmonic oscillations. In fact with a particular choice of coordinates the Lagrangian (\ref{lagr}) may be turned exactly into the canonical Lagrangian of the harmonic oscillator, but this is not necessary for our purposes. Solving the equations with the prescribed initial conditions $z(0)=q, \;\bar{z}(1)=\bar{y}$, we obtain $z(t)=q e^{2\im \alpha t},\;\bar{z}(t)=\bar{y}e^{-2\im \alpha (t-1)}$. Plugging this into the exponent of the path integral (\ref{evop}), we get $e^{-m\,\im \alpha}$. The term $\frac{1+z(1)\, \bar{y}}{1+q \bar{y}}$ in front of the exponent produces $\frac{1+q\, \bar{y} \,e^{2\im \alpha}}{1+q \bar{y}}$, and as a result we get
\bea\label{evopfin}
U(q,\bar{y})=\left(\frac{e^{-\im \alpha}+q\, \bar{y} \,e^{\im \alpha}}{1+q \bar{y}}\right)^m ,
\eea
which, as we know from (\ref{evopans}), is the right answer.

\comment{
In fact, the e.o.m. are easily linearized after the following change of variables:
\bea
z=\frac{v}{\sqrt{1+v\bar{v}}},\;\bar{z}=\frac{\bar{v}}{\sqrt{1+v\bar{v}}}
\eea
Then the measure takes the simple form $d\mu=dz\wedge d\bar{z}$ and the Lagrangian becomes quadratic:
\bea\label{lagr2}
\mathcal{L}=\frac{\im}{2}(\dot{\bar{z}}z-\dot{z}\bar{z})+\alpha(1-2\bar{z}z),
\eea
where in the first term we have performed an additional symmetrization using integration by parts. Clearly (\ref{lagr2}) describes a simple harmonic oscillator, and the e.o.m. are easily solved.
}

\section{Path integral for the spin chain}\label{pathint}

Similarly to what we did in (\ref{evop}), we now want to derive a path integral expression for the evolution operator of the spin chain $\hat{\mathbb{U}}=e^{i \alpha \hat{\mathbb{H}}}$, $\hat{\mathbb{H}}$ now being one of the Hamiltonians (\ref{aff})-(\ref{xxx}). Thus, we pass from the simple $\mathfrak{su}(2)$ case to the $\mathfrak{su}(3)$, or even $\mathfrak{su}(N+1)$ model. Before actually considering the Hamiltonians (\ref{aff})-(\ref{xxx}) we will start with a typical but simpler example of the $XXX$ Hamiltonian
\bea\label{xham}
H_{XXX}=\sum\limits_{i=1}^L\; P_{i,i+1}
\eea
and its generalizations to the symmetric powers of the fundamental representation, indexed by $m$ as before. These generalizations are obtained by replacing in (\ref{xham}) the permutation operator $P$ with $\pfat$, defined in (\ref{pfat}).

\subsection{The $XXX$ chain}\label{xxxchain}

In order to build the path integral we first need to know the matrix elements of the Hamiltonian itself, which essentially means that we need to know the matrix elements of the operator \( \pfat \). This operator acts in the tensor product \(\textrm{Sym}(\mathbf{C}^{N+1})^{\otimes m} \otimes \textrm{Sym}(\mathbf{C}^{N+1})^{\otimes m}\), and as an (overcomplete) basis in this space we will use the tensor product of the coherent state bases in each factor, i.e.
\bea
|\bar{a},\bar{b}\rangle = (1+v\circ \bar{a})^m\;(1+w\circ \bar{b})^m
\eea
Here and below \( \circ \) means a simple contraction (scalar product): \( x \circ \bar{y} = \sum\limits_{i=1}^{N}\,x_i \bar{y}_i \). The kernel of $\hat{\mathbb{P}}$ is easily found to be
\bea
\mathbb{P}(\bar{y}_1,\bar{y}_2 ; q_1, q_2)=\frac{ \langle\bar{q}_1,\bar{q}_2 | \hat{\mathbb{P}} | \bar{y}_1,\bar{y}_2\rangle }{\langle\bar{q}_1,\bar{q}_2  | \bar{y}_1,\bar{y}_2\rangle}=\left[\frac{ (1+q_1 \circ \bar{y}_2)\,(1+q_2\circ \bar{y}_1) }{ (1+ q_1\circ \bar{y}_1)\,(1+q_2\circ \bar{y}_2) }\right]^m
\eea
Now we can essentially repeat the steps from the previous Section. The only difficulty is notational and it comes from the fact that in this case, as opposed to the previous example, we essentially have two ``space-time'' directions: one ``time'' or $\alpha$-direction, and the second the direction, in which the spin chain is extended. As a consequence, our variables $z$ will now carry two indices: $z_{a,i}$, where $a$ is the time index, and $i$ enumerates the sites of the spin chain. The integrand will again split into two terms: one which may loosely be called the ``kinetic'' term and the second one being the Hamiltonian:
\bea
\mathbb{U}=  \underset{K\to\infty}{\textrm{lim}} \int\;\prod\limits_{a,i} d\mu(z_{a,i}, \bar{z}_{a,i})\; \times \mathbb{I}_{\mathrm{kin}} \times \mathbb{I}_{\mathrm{H}}
\eea

The kinetic term is local in the spin chain index $i$ and has a simplest (nearest-neighbor, or first-order) nonlocality, which is a general feature, since in the continuum limit it should lead to a 1-form:
\bea
\mathbb{I}_{\mathrm{kin}}=\prod\limits_{a,i} \left(\frac{1+z_{a,i}\circ \bar{z}_{a+1,i}}{1+z_{a+1,i}\circ \bar{z}_{a+1,i}}\right)^m
\eea
On the other hand, the Hamiltonian term has a first-order nonlocality in the spin-chain direction, but also has a first-order nonlocality in the time direction, since the matrix elements of the Hamiltonian entering the integral are always of the form
\bea\label{matel}
\frac{1}{K}\langle z_{a+1,i}|\widehat{\mathbb{H}}|z_{a,j}\rangle
\eea
This latter nonlocality will not play a role, since, as explicitly shown in (\ref{matel}), the contribution of such matrix element always comes with a damping factor $\frac{1}{K}$, and the nonlocality being of order $\frac{1}{K}$ as well enters only subleading terms. In any case, the contribution of the Hamiltonian may be written as
\bea
\mathbb{I}_{\mathrm{H}}=\prod\limits_{a,i} \left( 1+ m\,\frac{i \alpha}{K} \frac{1+ z_{a,i}\circ \bar{z}_{a+1,i+1}}{1+z_{a,i}\circ \bar{z}_{a+1,i}} \frac{1+z_{a,i+1}\circ \bar{z}_{a+1,i}}{1+z_{a,i+1}\circ \bar{z}_{a+1,i+1}} \right)
\eea
We may now exponentiate these expressions and take the limit $K \to \infty$, thus obtaining a continuous time variable $t$:
\bear\label{evopxxx}
\mathbb{U} = \int\; \prod\limits_{t\in [0,1]} d\mu(z(t),\bar{z}(t))\;\prod\limits_i \left(\frac{1+z_i(1)\, \bar{z}_i(1)}{1+z_i(0) \bar{z}_i(1)}\right)^m \;\exp{(\im\mathcal{S})}, \; \textrm{where}\\
\label{actionxxx}
\mathcal{S} = m\;\int\limits_0^1\,dt\, \sum\limits_i \left( i \frac{\dot{z}_i \circ \bar{z}_i}{1+z_i\circ \bar{z}_i} +
\alpha \frac{1+ z_{i}\circ  \bar{z}_{i+1}}{1+z_{i}\circ \bar{z}_{i}} \frac{1+z_{i+1}\circ \bar{z}_{i}}{1+z_{i+1}\circ \bar{z}_{i+1}} \right)
\eear
with boundary conditions $z_i(0)=q_i,\;\bar{z}_i(1)=\bar{y}_i$ . The nontrivial question is how to take the continuous limit in the spin chain direction, indexed by ``$i$'', --- there are several inequivalent ways to do it. It is well-known that the $XXX$ spin chain has two ``vacua'', i.e. the states with minimal and maximal energy. They also correspond to the extremal values of the spin: the vacuum with spin zero (or least possible spin in case the length of the chain does not allow for zero spin) is called antiferromagnetic, whereas the state with maximal spin (proportional to $L$ --- the length of the chain) is called ferromagnetic. Which one of these states is the true vacuum depends, of course, on the sign of the Hamiltonian.

\subsection{Ferromagnetic limit}\label{ferro}

The ferromagnetic limit is especially simple. It corresponds to the case where the $z$'s at the neighboring sites are very close to each other, that is $z_{i+1}-z_i \sim \frac{1}{L}$ (we remind the reader that $L$ is the length of the spin chain, i.e. the number of sites). The first term in (\ref{actionxxx}) then simply produces
\bea\label{ferrcont1}
 \int\,dt\, \sum\limits_i \left( \im \frac{\dot{z}_i \circ \bar{z}_i}{1+z_i \circ \bar{z}_i} \right) \to L \int \, dt\,\int\limits_{-1/2}^{1/2}\,dx\; \im \,\frac{\dot{z}(t,x) \circ \bar{z}(t,x)}{1+z(t,x)\circ \bar{z}(t,x)},
\eea
whereas the expression in the second term can be rewritten in the same spirit:
\bear
&&\frac{1+ z_{i}\circ  \bar{z}_{i+1}}{1+z_{i}\circ \bar{z}_{i}} \frac{1+z_{i+1}\circ \bar{z}_{i}}{1+z_{i+1}\circ \bar{z}_{i+1}}=\\ \nonumber
&&=1 -\sum\limits_i \frac{\Delta z_i \circ \Delta \bar{z}_i}{(1+z_i \circ \bar{z}_i)(1+z_{i+1} \circ \bar{z}_{i+1})}+{1\over 2} \sum\limits_i \frac{[(z_i \circ \Delta \bar{z}_i) \Delta z_i - (\Delta z_i \circ \Delta \bar{z}_i) z_i]\circ[\bar{z}_i+\bar{z}_{i+1}]}{(1+z_i \circ \bar{z}_i)(1+z_{i+1} \circ \bar{z}_{i+1})},
\eear
and in the continuum limit the last two terms reduce to
\bea\label{ferrcontlim}
{1\over L} \int dx\,\left(-\frac{\dx z \circ \dx \bar{z}}{(1+z \circ \bar{z})^2}+\frac{(z \circ \dx \bar{z}) (\dx z \circ \bar{z})-(z\circ \bar{z})(\dx z \circ \dx \bar{z})}{(1+z \circ \bar{z})^2}\right)=
-{1\over L} \int dx\,\left( \frac{\dx z \circ \dx \bar{z}}{1+z \circ \bar{z}} - \frac{(z \circ \dx \bar{z}) (\dx z \circ \bar{z})}{(1+z \circ \bar{z})^2} \right)
\eea
One immediately recognizes that the integrand in (\ref{ferrcontlim}) is the Fubini-Study metric (written in the inhomogeneous coordinates $z_{N+1}=\bar{z}_{N+1}=1$). Hence, the full action has the form
\bea\label{LL1}
\mathcal{S} = m\;\int\limits_0^1 \, dt\,\int\limits_{-1/2}^{1/2}\,dx\;\left[ L\, \im \,\frac{\dot{z}(t,x) \circ \bar{z}(t,x)}{1+z(t,x)\circ \bar{z}(t,x)} - {1\over L}\, \left( \frac{\dx z \circ \dx \bar{z}}{1+z \circ \bar{z}} - \frac{(z \circ \dx \bar{z}) (\dx z \circ \bar{z})}{(1+z \circ \bar{z})^2} \right) \right]
\eea
or after the rescaling $x\to \frac{1}{L} x $
\bea\label{LL}
\mathcal{S} = m\;\int\limits_0^1 \, dt\,\int\limits_\mathbb{R}\,dx\;\left[ \, \im \,\frac{\dot{z}(t,x) \circ \bar{z}(t,x)}{1+z(t,x)\circ \bar{z}(t,x)} - \, \left( \frac{\dx z \circ \dx \bar{z}}{1+z \circ \bar{z}} - \frac{(z \circ \dx \bar{z}) (\dx z \circ \bar{z})}{(1+z \circ \bar{z})^2} \right) \right]
\eea

Such non-relativistic sigma-models are known as Landau-Lifshitz models (see Appendix \ref{appLL}). The target space of the model we have described is, obviously, $\CP^N$. The simplest example corresponds to $N=1$, i.e. when the target space is a usual 2-sphere. In this case the model is also known as the classical Heisenberg ferromagnet, and it is customary to use the unit three-vector $\vec{n}$ instead of the complex coordinates $z, \bar{z}$ (the two parametrizations are related via the stereographic projection: $n_1+\im n_2=\frac{2z}{1+z\bar{z}}, \; n_3=\frac{1-z\bar{z}}{1+z\bar{z}}$). Then the e.o.m., which follows from Lagrangian (\ref{LL}), is:
\bea\label{heisferr}
\frac{\dd \vec{n}}{\dd t}=\vec{n}\times \frac{\dd^2 \vec{n}}{\dd x^2} .
\eea
Before concluding this Section let us return for a moment back to the expression (\ref{evopxxx}) and observe that, besides the action $\mathcal{S}$ in the exponent, it also includes a prefactor $\prod\limits_i \left(\frac{1+z_i(1)\, \bar{z}_i(1)}{1+z_i(0) \bar{z}_i(1)}\right)^m$, which in the continuum limit becomes
\bear
\prod\limits_i \left(\frac{1+z_i(1)\, \bar{z}_i(1)}{1+z_i(0) \bar{z}_i(1)}\right)^m
&\to&
\exp{\left[m\,L\,\int\limits_0^1 \,dx \, \left[\frac{1+z(t=1,x)\, \bar{z}(t=1,x)}{1+z(t=0,x) \bar{z}(t=1,x)}\right)\right]}
= \\ \nonumber &=&
\exp{\left[m\,L\,\int\limits_0^1 \,dx\, \left[\frac{1+z(t=1,x)\, \bar{y}(x)}{1+q(x) \bar{y}(x)}\right)\right]},
\eear
where in the last expression we took into account the boundary conditions. We will not need this expression in what follows, but it should not be overlooked.

\section{The antiferromagnetic limit}\label{anti}

The antiferromagnetic limit is much more difficult to analyze. The main idea is that in this case the $z$-variables on neighboring sites are no longer close to each other. Let us first elaborate on the case of the sphere, that is $N=1$, which was for the first time explored in \cite{H}. In this case it is intuitively clear that the antiferromagnetic limit corresponds to the case where the spins on the neighboring sites have opposite directions, i.e. $\vec{n}_{i+1}\simeq - \vec{n}_i$. In terms of the complex coordinates used above this may be written as $z_{i+1}\simeq -\frac{1}{\bar{z}_i}$, or, using homogeneous coordinates, as $z_1^{(i+1)} = \bar{z}_2^{(i)}, z_2^{(i+1)} = - \bar{z}_1^{(i)}$. Such simple explanation is due to the fact that on the sphere there exists the antipodal involution, that is the involution, which is an orientation-reversing isometry. The antipodal involution on the sphere is unique. On the other hand, this is no longer so for $\CP^N$ with $N \geq 2$.
%In particular, for even $N$ there are no involutions, and for odd $N$ there are infinitely many.
This is the reason why it is not immediately obvious, how one can extend the $\CP^1$ analysis to a higher-dimensional projective space. The answer crucially depends on the particular Hamiltonian at hand. The first model to be successfully analyzed was (\ref{aff}), so let us now recall how this was accomplished.

\subsection{The construction of Affleck}\label{aff1}

Affleck \cite{Affleck} considered a generalized Haldane limit for the spin chain with Hamiltonian (\ref{aff}). In order to rephrase his results in our language one should follow the steps of the previous Section to obtain the following action in the $t$-continuum limit:
\bea\label{actionaff}
\mathcal{S} = m\;\int\limits_0^1\,dt\, \sum\limits_i \left( i \frac{\dot{z}_i \circ \bar{z}_i}{z_i\circ \bar{z}_i} +
\alpha \frac{z_{i}\circ  z_{i+1}}{z_{i}\circ \bar{z}_{i}} \frac{\bar{z}_{i}\circ \bar{z}_{i+1}}{z_{i+1}\circ \bar{z}_{i+1}} \right)
\eea
The difference between the second terms in (\ref{actionxxx}) and (\ref{actionaff}) precisely reflects the difference between $\mathrm{P}$ and $\mathrm{Tr}$ operators entering the corresponding Hamiltonians. 
The minimum of the Hamiltonian $H=-\sum\limits_i  \frac{z_{i}\circ  z_{i+1}}{z_{i}\circ \bar{z}_{i}} \frac{\bar{z}_{i}\circ \bar{z}_{i+1}}{z_{i+1}\circ \bar{z}_{i+1}}$ is clearly reached for $z_{i+1}=\bar{z}_i$. The important observation is that for such configuration the first term in (\ref{actionaff}) turns into a full derivative, since on every two neighboring sites $i \frac{\dot{z}_i \circ \bar{z}_i}{z_i\circ \bar{z}_i}+i \frac{\dot{z}_{i+1} \circ \bar{z}_{i+1}}{z_{i+1}\circ \bar{z}_{i+1}}=i\frac{d}{dt}(\log{(z_i\circ \bar{z}_i)})$. There is a simple but fundamental explanation of this fact. In order to formulate it let us diverge for a moment to a slightly more general setup.

First of all, it is obvious that the direct product of two symplectic manifolds is a symplectic manifold. Indeed, let $\mathcal{M}_1, \mathcal{M}_2$  be endowed with the respective symplectic forms $\omega_1, \omega_2$. Then the natural symplectic form on $\mathcal{M}=\mathcal{M}_1 \times \mathcal{M}_2$ is $\omega=\omega_1 + \omega_2$. There's a canonical symplectic form on $\CP^{N}$, namely the Fubini-Study form, which looks as follows:
\bea\label{FS}
\omega_{FS}=\frac{da_i \wedge d\bar{a}_i}{\bar{a}_j a_j}-\frac{da_i \bar{a}_i \wedge d\bar{a}_k a_k}{(\bar{a}_j a_j)^2}.
\eea
The indices in this formula run from \(1\) to \(N+1\). According to the above remark, the symplectic form on the direct product of several $\CP^{N}$'s is the sum of the respective Fubini-Study forms. We may now formulate the following

\textbf{Observation 1.} The embedding of $\CP^N$ into $\CP^N\times \CP^N$, defined by the map $z\to(z,\bar{z})$, is Lagrangian.

Let us now expand the action (\ref{actionaff}) around the ``vacuum'' $z_{i+1}=\bar{z}_i$. The variables $z_{i+1}$ and $z_{i+2}$ are expressed in terms of $z_i$ in the following fashion:
\bea
z_{i+1}=\bar{z}_i+\ol \bar{\tau}_i,\qquad z_{i+2}=z_i +\ol z_i'
\eea
For convenience we introduce also the projector to the subspace of $\CC^{N+1}$ orthogonal to the vector $z_i$:
\bea
\Pi_i=\frac{\mathrm{I}}{|z_i|^2}-\frac{\bar{z}_i\otimes z_i}{|z_i|^4}
\eea

Then the terms in the Hamiltonian have the following expansions:
\begin{align}
\frac{|z_i \circ z_{i+1}|^2}{|z_i|^2 |z_{i+1}|^2}\simeq \olsq \tau_i \circ \Pi_i \circ \bar{\tau}_i,\quad
\frac{|z_{i+1} \circ z_{i+2}|^2}{|z_i|^2 |z_{i+1}|^2}\simeq \olsq \tilde{\tau}_i \circ \Pi_i \circ \overline{\tilde{\tau}}_i \quad \textrm{with}\quad \tilde{\tau}_i=\tau_i-z_i'
\end{align}

%%%%%%%%%%%%%%%%%%%%%%%%%%%%%%%%%%%%%%%%%%%%%%%%%%%%%%%%%%%%%%%%%%%%%%%%%
\comment{
\begin{align}
&\frac{|z_i \circ z_{i+1}|^2}{|z_i|^2 |z_{i+1}|^2}\simeq \olsq \left(\frac{\tau_i \circ \bar{\tau}_i}{|z_i|^2}-\frac{|\tau_i \circ \bar{z}_i|^2}{|z_i|^4} \right)\\
&\frac{|z_{i+1} \circ z_{i+2}|^2}{|z_i|^2 |z_{i+1}|^2}\simeq \olsq \left(\frac{\tilde{\tau}_i \circ \overline{\tilde{\tau}}_i}{|z_i|^2}-\frac{|\tilde{\tau}_i \circ \bar{z}_i|^2}{|z_i|^4} \right) \quad \textrm{with}\quad \tilde{\tau}_i=\tau_i-z_i'
\end{align}
}
%%%%%%%%%%%%%%%%%%%%%%%%%%%%%%%%%%%%%%%%%%%%%%%%%%%%%%%%%%%%%%%%%%%%%%%%%

The terms in the ``kinetic energy'' are expanded as follows:
\bea
i \frac{\dot{z}_i \circ \bar{z}_i}{z_i\circ \bar{z}_i}+i \frac{\dot{z}_{i+1} \circ \bar{z}_{i+1}}{z_{i+1}\circ \bar{z}_{i+1}}= i \ol \left[ \tau_i \circ \Pi_i \circ \dot{\bar{z}}_i - \bar{\tau}_i \circ \Pi_i \circ \dot{z}_i\right]
+ \textrm{full derivative}
\eea
Thus, the action (\ref{actionaff}) acquires the following form:
\bea
\mathcal{S} = m\;\int\limits_0^1\,dt\, \sum\limits_i \left( i \ol \left[ \tau_i \circ \Pi_i \circ \dot{\bar{z}}_i - \bar{\tau}_i \circ \Pi_i \circ \dot{z}_i\right] +\olsq \left[ \tau_i \circ \Pi_i \circ \bar{\tau}_i +(\tau_i-z_i') \circ \Pi_i \circ (\bar{\tau}_i-\bar{z}_i') \right]\right)
\eea
Now we simply need to ``integrate out'' the fields $\tau, \bar{\tau}$. Upon setting $\tau, \bar{\tau}$ equal to their stationary values we also pass to the continuum limit with respect to the ``$i$'' index. This leads to the following expression:
\bea\label{CPNaction}
\mathcal{S} = m\;\int\limits_0^1\,dt\,\int\limits_\mathbb{R}\,dx\;\left[{1\over 2} \partial_\mu z(x,t) \circ \Pi(z,\bar{z})\circ \partial_\mu \bar{z}(x,t)-{\im \over 2} \epsilon_{\mu\nu} \,\partial_\mu z(x,t) \circ \Pi(z,\bar{z})\circ \partial_\nu \bar{z}(x,t)  \right]
\eea
Clearly, the first term is the standard action of the $\CP^N$ sigma model, whereas the second term is the pull-back to the worldsheet of the K\"{a}hler form. The second term is topological and corresponds to the theta-angle $\theta = \pi m\;\;\textrm{mod}\;\;2\pi$.

\subsection{Antiferromagnetic configuration of the Hamiltonian (\ref{xxx})}\label{antifer}

We now want to move forward from the Hamiltonian (\ref{aff}) and find the sigma model which arises upon taking the continuum limit around the antiferromagnetic ``vacuum'' of the spin chain (\ref{xxx}). First of all, completely parallel to the discussion of the $XXX$ spin chain in Section \ref{xxxchain} above, we can write a path integral expression for the evolution operator of the spin chain (\ref{xxx}). Similarly to (\ref{actionxxx}), the action appearing in the exponent in the integrand of the path integral has the following form:
\bea\label{actionnew}
\mathcal{S} = m\;\int\limits_0^1\,dt\, \sum\limits_i \left( i \frac{\dot{z}_i \circ \bar{z}_i}{z_i\circ \bar{z}_i} +
\alpha \underbrace{\left(\frac{z_{i}\circ  \bar{z}_{i+1}}{z_{i}\circ \bar{z}_{i}} \frac{z_{i+1}\circ \bar{z}_{i}}{z_{i+1}\circ \bar{z}_{i+1}}+
 \frac{z_{i}\circ  \bar{z}_{i+2}}{z_{i}\circ \bar{z}_{i}} \frac{z_{i+2}\circ \bar{z}_{i}}{z_{i+2}\circ \bar{z}_{i+2}}\right)}_{\equiv \;\mathcal{H}}
 \right)
\eea
In this formula each of the variables $z_i$ has an additional (hidden) index, which takes three possible values corresponding to the fundamental representation of $SU(3)$. We emphasize that the Hamiltonian (\ref{xxx}) is interesting for us only in the case of $SU(3)$ symmetry --- in the $SU(N)$ case we pick a different Hamiltonian, see Section \ref{suN} below. We claim that in the case of (\ref{actionnew}) the antiferromagnetic vacuum configuration is when the $z$-vectors on any 3 neighboring sites are orthogonal to each other. First of all, this is consistent with what we had for the $SU(2)$ case above, since the equation $1+ \bar{z}_1 z_2=0$ arising in that case has a unique solution $z_2=-\frac{1}{\bar{z}_1}$, which is the antipodal involution discussed above. When $N=3$ we need to take three consecutive sites and impose orthogonality conditions on the three $z$-vectors $z_1, z_2, z_3$ sitting at these sites\footnote{Here we use homogeneous coordinates.}:
\bea\label{2flag}
z_1 \circ \bar{z}_2=z_2 \circ \bar{z}_3=z_1\circ \bar{z}_3=0 .
\eea
The submanifold of $(\CP^2)^{\times 3}$ described by (\ref{2flag}) is known as the flag manifold $\mathcal{F}_3$ (The index $3$ points out that this manifold is a homogeneous space of $SU(3)$. Flag manifolds for the group $SU(N)$ are introduced in Appendix \ref{appflag}.). We're now going to elaborate on this simplest nontrivial example.

\subsubsection{The $SU(3)$ case.}

Let us first reexamine the l.h.s. of (\ref{ferrcont1}). It is clear that the different continuum limit, described by (\ref{2flag}), will no longer produce the r.h.s. of (\ref{ferrcont1}). Thus, the question is what will arise in the continuum limit. The discussion above indicates that it is natural to first focus on arbitrary 3 consecutive sites. Then the kinetic term in the discretized Lagrangian is the pull-back $J_t$ of the following 1-form (hereafter we employ the homogeneous coordinates):
\bea\label{kinterm}
J= \im \frac{dz_1 \circ \bar{z}_1}{z_1 \circ \bar{z}_1}+\im \frac{dz_2 \circ \bar{z}_2}{z_2 \circ \bar{z}_2}+\im \frac{dz_3 \circ \bar{z}_3}{z_3 \circ \bar{z}_3}
\eea
This is the K\"{a}hler current on the product $\CP^2\times \CP^2 \times \CP^2$, and its divergence gives the K\"{a}hler (symplectic) form:
\bea
dJ=\Omega.
\eea
We claim that on the submanifold $\mathcal{F}_3$, described by (\ref{2flag}), this 2-form is zero. We may even formulate a slightly more general

\textbf{Observation 2.} The embedding $\cf_3 \hookrightarrow (\CP^2)^{\times 3}$ and more generally $\cf_{N} \hookrightarrow (\CP^{N-1})^{\times N}$ is Lagrangian.

The proof of this statement for $N=3$ is presented in Appendix \ref{appflag}, and for the moment let us focus on the consequences of this fact. It follows that
\bea
J|_{\mathcal{F}_3}=df,
\eea
$f$ being a function\footnote{In fact, $f=i \log{(\epsilon_{abc} \,z_1^a \,z_2^b \,z_3^c)}$.
%Indeed, let $z_1= \sqrt{z_1\circ \bar{z}_1}\, u_1,\;$ etc. and $U=\{u_1, u_2, u_3\}$ is a unitary matrix. Then $J_t=d({i\over 2} \log{(|z_1|^2 \, |z_2|^2 \,|z_3|^2)})+i \tr(dU U^\dagger)$
}, so the integral $\int\limits_0^1 J_t = f(1)-f(0)$ reduces to the boundary term. We ignore this term in the present discussion.

\section{The continuum limit}\label{cl}

The ``potential energy'' term $\mathcal{H}$ in (\ref{actionnew}) %, which in homogeneous coordinates takes the simple form
%\bea\label{hamperm}
%\mathcal{H}_i=\frac{z_{i}\circ  \bar{z}_{i+1}}{z_{i}\circ \bar{z}_{i}} \frac{z_{i+1}\circ \bar{z}_{i}}{z_{i+1}\circ \bar{z}_{i+1}}=\frac{|z_{i}\circ 
%\bar{z}_{i+1}|^2}{|z_{i}|^2 \; |z_{i+1}|^2},
%\eea
is equal to zero if we impose the background configuration (\ref{2flag}): $z_{i}\circ  \bar{z}_{i+1}=0$. Moreover, since $0\leq\mathcal{H}\leq 2$, one immediately sees that the ferromagnetic and antiferromagnetic vacua saturate respectively the maximum and minimum of its possible values. In view of the fact that in the following we will build an expansion around the antiferromagnetic vacuum, from this observation we deduce an important consequence, namely that this expansion must start with a quadratic term (at least), i.e. there is no linear term.

Let us assume that the number of sites of our spin chain is a factor of 3 (this is only needed for simplicity, and it does not play a big role for a sufficiently long spin chain). In this case we split the spin chain into $\hat{L}$ segments of length $3$ and focus for the moment on just one of these segments, which is link number $k$ in the chain.

\subsection{The expansion around the ``vacuum'' configuration}\label{expansion}

On each of the three sites we have a three-dimensional complex vector $z$. Let us form a $3\times 3$ matrix of these vectors, which we denote by $Z$.
%Each of these vectors is defined up to multiplication by an arbitrary complex number, which means that the matrix is defined up to right multiplication by a nondegenerate diagonal matrix. We will denote the corresponding quotient by $\hat{\mathbb{P}}(\mathrm{Mat}_3)$.
The antiferromagnetic configuration corresponds to the case where the three vectors are mutually orthogonal.
%, which means that in $\hat{\mathbb{P}}(\mathrm{Mat}_3)$ we may represent $Z$ by a unitary matrix.
Now we need to take the fluctuations into account, and in order to build the sought for expansion we will employ the so-called $QR$ decomposition of a matrix. The $QR$ decomposition theorem says that an arbitrary matrix $Z$ may be decomposed into a product of a unitary matrix $U$ and an upper triangular one $B_+$:
\bea\label{QR}
Z= U\circ B_+
\eea
This statement is equivalent to the Gram-Schmidt orthogonalization theorem. Let us parametrize $B_+$ in the following way:
\bea
B_+= \begin{pmatrix} 
  1 & \ol \xs_k & \ol\ys_k \\ 
  0 & 1 & \ol\zs_k \\
  0 & 0 & 1 \\
\end{pmatrix} \begin{pmatrix} 
  \as_k & 0 & 0 \\ 
  0 & \bs_k & 0 \\
  0 & 0 & \cs_k \\
\end{pmatrix} 
\eea
%The difference of the case under consideration from the generic situation is that here we work in the projectivization $\hat{\mathbb{P}}(\mathrm{Mat}_3)$, which means that we may safely set $\as=\bs=\cs=1$.
If we denote the columns of the matrix $U$ as $(u_1, u_2, u_3)$, the decomposition (\ref{QR}) says that
\bea\label{contin2}
z_{1,k} = \as_{k}\, u_{1,k},\quad z_{2,k} = \bs_k\,(u_{2,k}+ \ol\xs_k\, u_{1,k}),\quad z_{3,k}=\cs_k \,(u_{3,k}+\ol\ys_k\, u_{1,k}+ \ol\zs_k\, u_{2,k})
\eea
The hypothesis of the existence of a continuum limit implies that $u_{1,k},\;u_{2,k},\;u_{3,k}$ vary mildly with $k$, in other words we may approximate
\bea\label{contin}
u_{i,m+1}=u_{i,m}+\ol u_{i,m}'+ ...
\eea

%%%%% Here comes the picture Fig.1 %%%%%
\begin{pspicture}(15,5)
%\psset{linecolor=blue} 
\pnode(0.5,3){e1}
\cnode*(1.5,3){.2}{a1}
\cnode*(3,3){.2}{a2}
\cnode*(4.5,3){.2}{a3}
\cnode*(6,3){.2}{a4}
\cnode*(7.5,3){.2}{a5}
\cnode*(9,3){.2}{a6}
\cnode*(10.5,3){.2}{a7}
\cnode*(12,3){.2}{a8}
%\cnode*(13.5,3){.2}{a9}
\pnode(13,3){e2}
\rput(3,3){\footnotesize \color{white} $1$}
\rput(4.5,3){\footnotesize \color{white} $2$}
\rput(6,3){\footnotesize \color{white} $3$}
\rput(7.5,3){\footnotesize \color{white} $1$}
\rput(9,3){\footnotesize \color{white} $2$}
\rput(10.5,3){\footnotesize \color{white} $3$}
\psset{linestyle=dashed}
\ncline{e1}{a1}
\ncline{a1}{a2}
%\ncput*{$H_1$}
\ncline{a2}{a3}
\ncline{a3}{a4}
\ncline{a4}{a5}
\ncline{a5}{a6}
\ncline{a6}{a7}
\ncline{a7}{a8}
%\ncline{a8}{a9}
\ncline{a8}{e2}
\psset{linestyle=solid}
\ncbar[offsetB=-2pt,angleA=90,nodesep=3pt,armB=20pt]{-}{a3}{a5}
\naput[labelsep=2pt]{\footnotesize $\mathcal{H}_{k-1,k}^{2,1}$}
\ncbar[offsetB=2pt,angleA=-90,nodesep=3pt]{-}{a4}{a5}
\nbput[labelsep=2pt]{\footnotesize $\mathcal{H}_{k-1,k}^{3,1}$}
\ncbar[offsetB=2pt,offsetB=-2pt,angleA=90,nodesep=3pt,armB=3pt,armA=3pt]{-}{a4}{a6}
\naput[npos=1.75,labelsep=2pt]{\footnotesize $\mathcal{H}_{k-1,k}^{3,2}$}
\ncbar[offsetA=2pt,angleA=-90,nodesep=3pt,armB=3pt,armA=3pt]{-}{a5}{a6}
\nbput[labelsep=1pt]{\footnotesize $\mathcal{H}_{k,k}^{1,2}$}
\ncbar[offsetA=0pt,angleA=-90,nodesep=3pt,armB=20pt,armA=20pt]{-}{a5}{a7}
\nbput[labelsep=2pt]{\footnotesize $\mathcal{H}_{k,k}^{1,3}$}
\ncbar[offsetA=-2pt,angleA=90,nodesep=3pt]{-}{a6}{a7}
\naput[labelsep=2pt]{\footnotesize $\mathcal{H}_{k,k}^{2,3}$}
%\psset{linecolor=red} 
\ncbox[nodesep=.2cm,boxsize=.3,linearc=.2]{<->}{a2}{a4}
\nbput[npos=0.5]{\footnotesize $k-1$}
\ncbox[nodesep=.2cm,boxsize=.3,linearc=.2]{<->}{a5}{a7}
\nbput[npos=0.7]{\footnotesize $k$}
\rput(7,1){Fig. 1. Explanation of the various terms calculated in (\ref{hamcontlimit}).}
\end{pspicture}
%%%%%%%%%%%%%%%%%%%%%%%%

Let us introduce the quantity
\bea\label{hamperm2}
\mathcal{H}_{m,n}^{i,j}=\frac{|z_{i,m}\circ  \bar{z}_{j,n}|^2}{|z_{i,m}|^2 \; |z_{j,n}|^2},
\eea
which is the density of the Hamiltonian $\mathcal{H}$ from (\ref{actionnew}), if the indices $i, j, m, n$ change in a particular range. Indeed, we need to calculate $\mathcal{H}_{m,n}^{i,j}$ for nearest- and next-to-nearest neighbor sites,
%generalizing (\ref{hamperm}),
using the formulas (\ref{contin2})-(\ref{contin}) and keeping only the leading orders of $\olsq$ (see Fig.1 for an explanation of what these terms stand for):
\begin{align} \nonumber
&\mathcal{H}_{k,k}^{1,2}=\frac{|z_{1}\circ  \bar{z}_{2}|^2}{|z_{1}|^2 \; |z_{2}|^2}\simeq \olsq |\xs_k|^2,\quad\quad
\mathcal{H}_{k,k}^{2,3}=\frac{|z_{2}\circ  \bar{z}_{3}|^2}{|z_{2}|^2 \; |z_{3}|^2}\simeq \olsq |\zs_k|^2,\quad\quad
\mathcal{H}_{k,k}^{1,3}=\frac{|z_{1}\circ  \bar{z}_{3}|^2}{|z_{1}|^2 \; |z_{3}|^2}\simeq \olsq |\ys_k|^2 \displaybreak[3]\\ \label{hamcontlimit}
&\mathcal{H}_{k-1,k}^{3,1}=\frac{|z_{3,k-1}\circ  \bar{z}_{1,k}|^2}{|z_{3,k-1}|^2 \; |z_{1,k}|^2}
=\frac {|(u_{3,k-1}+\ol \ys_{k-1} \, u_{1,k-1}+\ol \zs_{k-1} u_{2,k-1})\circ \bar{u}_{1,k} |^2}{|u_{3,k-1}+\ol \ys_{k-1} \, u_{1,k-1}+\ol \zs_{k-1} u_{2,k-1}|^2 |u_{1,k}|^2}\simeq
\\ \nonumber &~~~~~~~~~~\simeq
\olsq |-u_{3,k}' \circ \bar{u}_{1,k}+\ys_{k}|^2 \displaybreak[3] \\ \nonumber
&\mathcal{H}_{k-1,k}^{3,2}=\frac{|z_{3,k-1}\circ  \bar{z}_{2,k}|^2}{|z_{3,k-1}|^2 \; |z_{2,k}|^2}
=\frac {|(u_{3,k-1}+\ol \ys_{k-1} \, u_{1,k-1}+\ol \zs_{k-1} u_{2,k-1})\circ (\bar{u}_{2,k}+ \ol\bar{\xs}_k\, \bar{u}_{1,k}) |^2}{|u_{3,k-1}+\ol \ys_{k-1} \, u_{1,k-1}+\ol \zs_{k-1} u_{2,k-1}|^2 |u_{2,k}+ \ol\xs_k\, u_{1,k}|^2}\simeq\\ \nonumber
&~~~~~~~~~~\simeq \olsq |-u_{3,k}' \circ \bar{u}_{2,k}+\zs_{k}|^2 \displaybreak[3]\\ \nonumber
&\mathcal{H}_{k-1,k}^{2,1}=\frac{|z_{2,k-1}\circ  \bar{z}_{1,k}|^2}{|z_{2,k-1}|^2 \; |z_{1,k}|^2}
=\frac {|(u_{2,k-1}+ \ol\xs_{k-1}\, u_{1,k-1})\circ \bar{u}_{1,k} |^2}{|u_{2,k-1}+ \ol\xs_{k-1}\, u_{1,k-1}|^2 |u_{1,k}|^2}\simeq \olsq |-u_{2,k}' \circ \bar{u}_{1,k}+\xs_{k}|^2
\end{align}
%\mathcal{H}_{k,k+1}^{3,1}=\frac{|z_{2}\circ  \bar{z}_{3}|^2}{|z_{2}|^2 \; |z_{3}|^2}\simeq \olsq |z|^2\\
%\mathcal{H}_{k,k+1}^{2,1}=\frac{|z_{2}\circ  \bar{z}_{3}|^2}{|z_{2}|^2 \; |z_{3}|^2}\simeq \olsq |z|^2\\
%\mathcal{H}_{k,k+1}^{3,2}=\frac{|z_{2}\circ  \bar{z}_{3}|^2}{|z_{2}|^2 \; |z_{3}|^2}\simeq \olsq |z|^2\\

Hence, the Hamiltonian $H_2=\sum\limits_i (P_{i,i+1}+ P_{i,i+2})$ produces a contribution
\bea\label{hamflag}
\mathcal{H}=\olsq \!\!\left[ \sum\limits_k  \left(|\xs_k|^2+\!|\ys_k|^2\!+\!|\zs_k|^2+\!|-u_{3,k}' \circ \bar{u}_{1,k}+\ys_{k}|^2\!+\!|-u_{3,k}' \circ \bar{u}_{2,k}+\zs_{k}|^2\!+\!|-u_{2,k}' \circ \bar{u}_{1,k}+\xs_{k}|^2 \right)
\right]
\eea

%%%%%%%%%%%%%%%%%%%%%%%%%%%%%%%%%%%%%%%%%%%%%%%%%%%%%%%%%%%%%%%%%%%%%%
\comment{
Hence, the Hamiltonian $H=J_1 \sum\limits_i P_{i,i+1}+ J_2 \sum\limits_i P_{i,i+2}$ produces a contribution
\bea
\mathcal{H}=\olsq \left[J_1 \sum\limits_k  \left(|\xs_k|^2+|\zs_k|^2+|-u_{3,k}' \circ u_{1,k}+y_{k}|^2 \right)+J_2 \sum\limits_k \left(|y_k|^2+|-u_{3,k}' \circ u_{2,k}+z_{k}|^2+|-u_{2,k}' \circ u_{1,k}+x_{k}|^2 \right)
\right]
\eea
}
%%%%%%%%%%%%%%%%%%%%%%%%%%%%%%%%%%%%%%%%%%%%%%%%%%%%%%%%%%%%%%%%%%%%%%

Let us now turn to the kinetic term (\ref{kinterm}) and see what it produces to the leading order in $\ol$. A simple calculation reveals that
\bear\label{kinetflag}
J_t =
%\im (\dot{u_1} \bar{u}_1-u_1 \dot{\bar{u}}_1+\dot{u_2} \bar{u}_2-u_2 \dot{\bar{u}}_2+\dot{u_3} \bar{u}_3-u_3 \dot{\bar{u}}_3)
\left( \im \frac{\dot{z}_1 \circ \bar{z}_1}{z_1 \circ \bar{z}_1}+\im \frac{\dot{z}_2 \circ \bar{z}_2}{z_2 \circ \bar{z}_2}+\im \frac{\dot{z}_3 \circ \bar{z}_3}{z_3 \circ \bar{z}_3}\right)_{\xs=\ys=\zs=0} \!\!\!\!-\;\;%\\ \nonumber
\frac{\im}{L} \left( \,\xs\; u_1 \circ \dot{\bar{u}}_2 + \, \ys\;  u_1 \circ \dot{\bar{u}}_3 + \zs\; u_2 \circ \dot{\bar{u}}_3  - \textrm{c.c.}\right)\;+ ...
% (\dot{u}_1 \bar{u}_2 - 
%(\dot{u_1}\bar{u}_3
%\dot{u_2} \bar{u}_3
\eear
%where the $v$'s represent the background configuration for the $z$ fields with the fluctuations set to zero: $\xs=\ys=\zs=0$ (in other words, $v_1 = \as \,u_1,\;v_2= \bs \,u_2,\; v_3=\cs \,u_3$).
But we have proved above that the first line of this expression is in fact a full derivative:
\bea
\left( \im \frac{\dot{z}_1 \circ \bar{z}_1}{z_1 \circ \bar{z}_1}+\im \frac{\dot{z}_2 \circ \bar{z}_2}{z_2 \circ \bar{z}_2}+\im \frac{\dot{z}_3 \circ \bar{z}_3}{z_3 \circ \bar{z}_3}\right)_{\xs=\ys=\zs=0} =\;\;\frac{\textrm{d} f}{\textrm{d} t}
\eea
and, as such, can be omitted up to boundary terms. Next we combine (\ref{hamflag}) and (\ref{kinetflag}) and ``integrate out'' the auxiliary variables $\xs_k, \ys_k, \zs_k$, whereupon we obtain:
\begin{align}\label{flagaction1}
\mathcal{S}=&\frac{1}{2}\,\int dt \, dx\; \left( \left[ |u_1 \circ \partial_\mu \bar{u}_2|^2+|u_1 \circ \partial_\mu \bar{u}_3|^2+|u_2 \circ \partial_\mu \bar{u}_3|^2 \right]
+ \right. \\ \nonumber 
&\left. +\,
\im \,\epsilon_{\mu\nu} \left[ (u_1 \circ \partial_\mu \bar{u}_2)\;( \bar{u}_1 \circ \partial_\nu u_2)+ (u_1 \circ \partial_\mu \bar{u}_3)\;( \bar{u}_1 \circ \partial_\nu u_3)+  (u_2 \circ \partial_\mu \bar{u}_3)\;( \bar{u}_2 \circ \partial_\nu u_3) \right] \right).
\end{align}
The first line in (\ref{flagaction1}) is the so-called normal metric on the flag space $\mathcal{F}_3$ (see Section \ref{flagmetr}), whereas the second line is the pull-back to the worldsheet of $\Omega|_\mathcal{F}$, where $\Omega|_\mathcal{F}$ is the restriction to the flag of the symplectic form $\Omega$ on $(\CP^2)^{\times 3}$. However, as discussed above, this restriction is identically zero, so the second line vanishes, and we are left with
\bea\label{flagaction}
\mathcal{S}=\frac{1}{2}\,\int dt \, dx\; \left(  |u_1 \circ \partial_\mu \bar{u}_2|^2+|u_1 \circ \partial_\mu \bar{u}_3|^2+|u_2 \circ \partial_\mu \bar{u}_3|^2 \right),
\eea
and one should keep in mind that $u_{1,2,3}$ are subject to the orthonormality conditions. A rather interesting property of the action (\ref{flagaction}), which is in contrast to the action (\ref{CPNaction}) above, is that it does not contain a $\theta$-term, although such a term is not prohibited by any symmetries. Moreover, there's actually space for two $\theta$-angles, since the corresponding cohomology group $H^2(\cf_3, \mathbb{R})\simeq \mathbb{R}^2$ is two-dimensional\footnote{There is in fact a simple choice for the de-Rham representatives of this cohomology group. Denote by $\tilde{\Omega}_i = \Omega_i|_{\cf_3},\; i=1,2,3$ the restrictions to the flag manifold of the three Fubini-Study forms. The form $\tilde{\Omega}=\sum\limits_{i=1}^3\;\theta_i\, \tilde{\Omega}_i$ is closed. Since, as discussed above, $\sum\limits_i\; \tilde{\Omega}_i=0$, the form $\tilde{\Omega}$ is parametrized by a vector $(\theta_1, \theta_2, \theta_3)\;\;\textrm{mod}\;\; (1, 1, 1)$, which therefore determines a two-dimensional space. As one can check, it is the space of non-exact closed two-forms.}.

\subsection{Metrics on flag manifolds}\label{flagmetr}

In this Section we will show that the metric (\ref{flagaction}) on $\mathcal{F}_3$ that we have obtained is not the most general metric compatible with the symmetries of the flag manifold\footnote{I am grateful to A.Gerasimov and S.Shatashvili for pointing out to me that there is a family of metrics on $\mathcal{F}_N$ .}. This is rather obvious from the beginning, since it is clear that we can multiply each of the three terms in (\ref{flagaction}) by an arbitrary (positive) constant without breaking the $SU(3)$ symmetry. Below we present the most general $SU(N)$-invariant metric on $\cf_N$, but in many cases one is interested in the special metrics possessing certain properties, such as being K\"{a}hler or Einstein --- the reader can find a rather detailed discussion of these properties in \cite{greek}. In particular, it follows from this work that the normal metric, entering the action (\ref{flagaction}) above, is Einstein but not K\"{a}hler (although the flag manifold is a K\"{a}hler manifold meaning that it can be equipped with a K\"{a}hler metric).

Let us start from the obvious fact that, since $U(1)^N \subset U(N)$ the Lie algebra $\ug(N)$ naturally splits
\bea
\ug(n)=\underset{n \;\textrm{times}}{\underbrace{\ug(1)\oplus ... \oplus \ug(1)}}\oplus \hg \equiv \tgo \oplus \hg
\eea
This is in fact nothing but the decomposition of a Lie algebra into a Cartan subalgebra and the associated root space $\hg= \underset{\textrm{all roots}\;\alpha}{\bigoplus} E_\alpha$, $E_\alpha$ being the subspace corresponding to a given root. $\mathcal{F}_3$ is not a symmetric space, i.e. $[\hg, \hg] \nsubseteq \tgo $ --- this is obvious, since the commutator of two ``roots'' in general produces another root\footnote{With the exception of $\mathfrak{su}(2)$ where there's only one positive and negative root. In this case $\mathcal{F}_2\simeq SU(2)/U(1)\simeq S^2$, i.e. a sphere, which clearly \emph{is} a symmetric space.} $[E_\alpha, E_\beta]\subset E_{\alpha+\beta}$ and not a Cartan generator.

Next we review the standard ``coset'' construction of metrics. One generally takes a group element $g\in U(N)$ and builds a current
\bea
J=g^{-1}dg \;\in\; \ug(N) .
\eea
The action of the stabilizer $T=U(1)^N$ of the coset on $J$ is as follows:
\bea
h \circ J = (g h)^{-1} d(gh)=h^{-1} J h + h^{-1} dh
\eea
The last term in this expression belongs to $\tgo$. Thus, the transformation of $\pi(J)$ --- the projection of $J$ on $\hg\subset \ug(N)$ --- is particularly simple:
\bea
h \circ \pi(J) = h^{-1} \pi(J) h,
\eea
or in other words $\tgo$ is represented on $\pi(J)$. One is often accustomed to writing the metric on a quotient space $G/T$ (with Lie algebra splitting $\mathfrak{g}=\hg\oplus \tgo$) simply as $\tr(\pi(J)^2)$, but an important fact to realize is that this is only unique, when the representation of $T$ on $\hg$ is \emph{irreducible}. This is not so for the case at hand (and indeed an irreducible (over $\CC$) representation of $U(1)$ must be one-dimensional --- this is also true in our situation, since the $N$ groups $U(1)$ commute).

It is easy to see that each component $j_{ab},\;a\neq b$ of the matrix $J$ furnishes an irreducible representation of $U(1)^N$ with the weight 
\bea
(0,\;...\;, \underset{\underset{\textrm{a-th place}}{\uparrow}}{1},\;...\;, \underset{\underset{\textrm{b-th place}}{\uparrow}}{-1},\;...\;,0),
\eea
which means that for all $a> b$ the quantity $j_{ab} j_{ba}=|j_{ab}|^2$ is an invariant. This means precisely that the generic metric can be written as follows:
\bea
ds^2 = \sum\limits_{a>b} C_{ab} |j_{ab}|^2\quad\textrm{with}\quad C_{ab} >0,
\eea
the latter requirement needed for the nondegeneracy of the metric.

%\section{Lagrangian embeddings}

%Let $(N,\Omega)$ be a symplectic manifold, and $M$ its Lagrangian submanifold: $\Omega|%_M =0$.

\section{Arbitrary $N$, the mass gap and relation to trimerization ($N$-merization)}\label{suN}

So far we have mainly concentrated on the case of a spin chain (\ref{xxx}) for the symmetry group $SU(3)$. It can be easily generalized to the case of the group $SU(N+1)$. To do it one needs to replace (\ref{xxx}) by the following Hamiltonian:
\bea
H_{N}= \sum\limits_{i=1}^{i=L} \, \sum\limits_{k=1}^{N}\, P_{i,i+k} .
\eea
Repeating the manipulations explained in the previous Sections, one arrives at the sigma model for the flag manifold $\cf_{N+1}$ with the $S_{N+1}$-symmetric choice of metric.

We now pass to a particular application of the general theory. Recently there has been interest in the spin chain models exhibiting the phenomenon of ``trimerization'', or even more generally ``$N$-merization''. This means that the $N$ neighboring sites on a spin chain form a bond --- in other words, the $N$ fundamental representations of $SU(N)$ combine into a singlet. This could well be called ``baryonization'', if one prefers the elementary particle terminology. When $N=2$, the phenomenon is known as dimerization and has been well studied (see \cite{RS}, for instance). There is even a model, which exhibits exact dimerization of the ground state --- the so-called Majumdar-Ghosh model, and recently generalizations to the $N$-merized case have been introduced \cite{greiter}. It follows from the results of this paper that the Hamiltonian (\ref{xxx}) exhibits trimerized order.

The authors of \cite{corboz} considered a spin chain, which is directly relevant to our discussion in this paper. It is described by the following Hamiltonian:
\bea\label{corbozham}
\tilde{H}=J_1\;\sum\limits_i\; \left(\cos{\theta}\;(\mathbf{S}_i \mathbf{S}_{i+1}) +\sin{\theta} \;(\mathbf{S}_i \mathbf{S}_{i+1})^2\right)
+ J_2\;\sum\limits_i\; \left(\cos{\theta}\;(\mathbf{S}_i \mathbf{S}_{i+2}) +\sin{\theta} \;(\mathbf{S}_i \mathbf{S}_{i+2})^2\right)
\eea
Here $\mathbf{S}$ are the $su(2)$ spin operators written in the vector $\mathbf{3}$ representation.

One can check that the operator $P=(\mathbf{S}_i \mathbf{S}_{i+1}) + (\mathbf{S}_i \mathbf{S}_{i+1})^2$ is simply a permutation in the tensor product $\CC^3 \otimes \CC^3$. It means that for $J_2 = J_1\equiv J$ and $\theta={\pi \over 4}$ the above $SU(2)$-invariant Hamiltonian reduces to the $SU(3)$-invariant $\tilde{H}= J\,\sum\limits_i\; (P_{i,i+1}+P_{i,i+2})$, which is nothing but our Hamiltonian (\ref{xxx}). Figure 2 reproduces a picture from the paper \cite{corboz}, representing the phase diagram of the model (\ref{corbozham}), which was obtained numerically.
\begin{figure}[htb]
\renewcommand{\figurename}{}
\begin{center}
\includegraphics[scale=0.4]{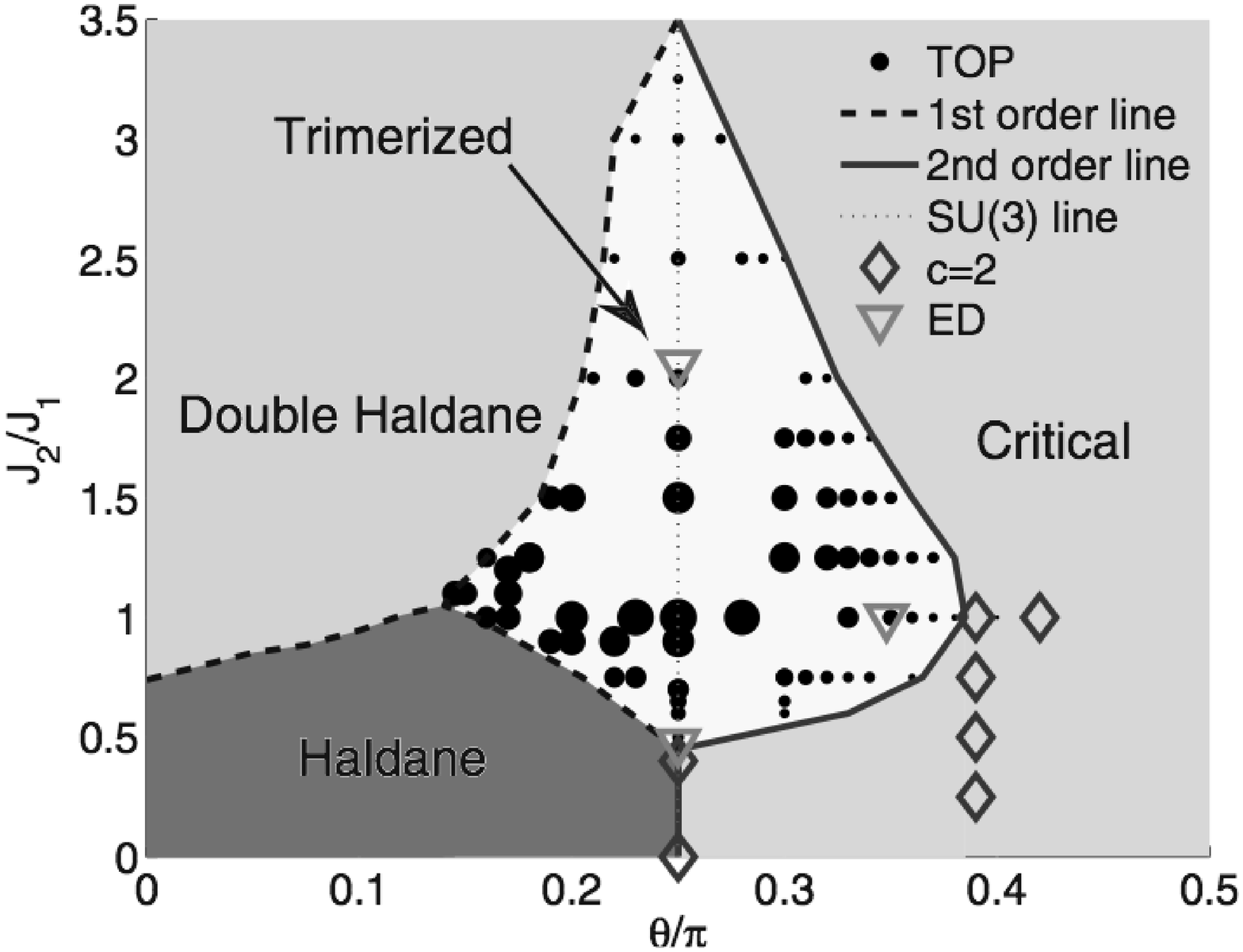}
\end{center}
\caption*{Fig. 2. Phase diagram of the Hamiltonian (\ref{corbozham}) \cite{corboz}.} 
\end{figure}
In particular, one can see that for the prescribed values of the parameters we get directly into the center of the trimerized phase. The numerical calculations performed by the authors of \cite{corboz} also suggest that, similarly to the Haldane gap of the integer spin $\vec{S}_i \vec{S}_{i+1}$ chain, the trimerized phase has a gap, too. It would be very interesting to show this analytically for the flag sigma model described by the action (\ref{flagaction}), but so far we have not been able to advance in this direction. The reason for this is that the usual method for this kind of calculations --- the $1/N$ expansion --- does not work in this case, since the model (\ref{flagaction}) is more of a matrix model than a vector model. Indeed, the dynamical degrees of freedom are the $N$ vectors with $N$ components each. Essentially the problems we encounter along the $1/N$ expansion route are similar to the ones that arise in the principal chiral model \cite{Polyakov}.
%\section{The integrable models}
%\subsection{$\CP^N$ with fermions}

\section{Discussion}\label{disc}

In this paper we found a spin chain, whose excitations near the antiferromagnetic vacuum are described by the sigma-model with target space the manifold of complete flags\footnote{$B$ here stands for the Borel subgroup of $GL(3,\CC)$.} $\cf_3\simeq GL(3,\CC)/B \simeq U(3)/U(1)^3$ (with a special $S_3$-symmetric choice of metric). We believe that the flag sigma model captures the low-energy dynamics over the trimerized vacuum configuration (or over the $N$-merized vacuum in the case of $SU(N)$) of a particular spin chain. This seems to be consistent with the recent results in condensed matter \cite{corboz,greiter}. 
%We have also shown that the the spectrum of excitations has a mass gap in this model.

Besides these applications, we hope our approach can be generalized to other situations. In particular, it opens the possibility of searching for various Lagrangian embeddings and building the corresponding spin chains. It is also worth emphasizing, that although in this paper we were dealing solely with the Lagrangian embeddings, this is in fact too strong a requirement, and in principle an isotropic embedding is sufficient for our argumentation (An embedding $M\hookrightarrow (N, \Omega)$ is said to be isotropic, if $\Omega|_M=0$ --- that is, no restriction on the dimension of $M$ is imposed). Such situations have been widely explored in the symplectic geometry literature \cite{weinstein} and it would be interesting to understand, what they mean in the spin chain setup. 

Another interesting question has to do with the integrability properties of the spin chains/sigma models. As explained in the introduction, this was actually the main motivation of our work --- to establish integrability/non-integrability of various $\CP^N$ models with fermions (model (\ref{final}) in particular). It has been known for a while now that the bosonic $\CP^N$ model is not integrable, and its ``would-be'' solitons are confined \cite{adda}. On the other hand, once you add the fermions, the situations can change dramatically, and the solitons can in some cases become liberated \cite{koberle}.

As we discussed above, the $\CP^N$ sigma model has been obtained from an alternating spin chain in \cite{Affleck}. It is also likely that the fermionic $\CP^N$ models can be obtained in a similar spirit (target-space supersymmetric $\CP^{N|M}$ models have been obtained in \cite{saleur}). The main difficulty, however, is that in general these spin chains for arbitrary values of representation (or spin) $S$ are not integrable\footnote{I am grateful to F.Smirnov for drawing my attention to this fact.}. Thus, the question splits into two:
\begin{itemize}
\item 1) whether one can find a family (parametrized by $S$) of integrable spin chains, giving rise to the desired sigma model in the continuum limit \textbf{and}
\item 2) if not, how one can find traces of integrability properties of the continuum sigma model in the spin chain (even if the spin chain itself is not integrable).
\end{itemize}
In our opinion, it would be very interesting to carry out this program for the $\CP^N$ models, in particular to see on the level of the spin chain why the bosonic $\CP^N$ model and the $\CP^N$ model with fermions differ so much.

\vspace{0.75cm}

\textbf{Acknowledgments.} I am grateful to Sergey Frolov for numerous useful and illuminating discussions in the course of work and for carefully reading the manuscript. I am also indebted to Anton Gerasimov, Samson Shatashvili, Fedor Smirnov and Kostya Zarembo for several valuable comments that I have benefitted from. My work was supported by the Irish Research Council for Science, Engineering and Technology, in part by grants RFBR 11-01-00296-a
%, 09-01-12150-ofi\_m
and in part by grant for the Support of Leading Scientific Schools of Russia NSh-8265.2010.1.

\appendix
\section{Definitions.}\label{app1}

 Let $V$ be a vector space. Its tensor square is denoted by $W=V\otimes V$. The permutation operator $P\in \mathrm{End}(W)$ is defined in an obvious way: $P(a\otimes b)=b\otimes a$. The definition of the trace operator $\mathrm{Tr}$ is slightly less obvious. Let $s$ be  the ``scalar product'' in $V$, in other words a $\CC$-linear nondegenerate symmetric operator $s: W\to \CC$ \footnote{Note that this convention is slightly different from the usual one, when $s$ is taken to be semilinear. This is done so that we can realize the trace operator as a matrix acting on vectors, which is in the spirit of the spin chain.}. We also fix the scalar product on $W$: $(a\otimes b, c\otimes d)=(a, c)(b, d)=s(a\otimes c) s(b\otimes d)$. Since $s\in W^\ast$, using the scalar product on $W$ we find a unique dual element $s^\ast \in W$. In more detail, its defining property is $(s^\ast,c\otimes d)=s(c\otimes d)=(c,d)$. The trace operator, which is an element of $\mathrm{End}(W)\simeq W \otimes W^\ast$, is defined as follows: $\mathrm{Tr}=s^\ast \otimes s$. This means that $\mathrm{Tr} (a\otimes b)=(a,b) s^\ast$ and, for instance, $(\mathrm{Tr} (a\otimes b), c\otimes d)=(a,b)(c,d)$.

\section{Landau-Lifshitz models.}\label{appLL}

Landau-Lifshitz models are Hamiltonian systems of a particular type\footnote{In this Section we mainly follow the exposition of \cite{AH}, slightly adapting it to our needs.}. The main dynamical variable is a function $m(x)$, taking values in a Lie algebra $\mathfrak{g}$ (corresponding to a Lie group which we denote by $G$). The variable $x$ may take values in $\mathbb{R}$, although a more complete mathematical theory exists for the case where $x\in S^1$. The Poisson structure is defined as follows:
\bea\label{poisson}
\{m^a(x), m^b(y)\}=\im \, f^{ab}_c \,\delta(x-y) \,m^{c}(x),
\eea
where $f^{ab}_c$ are the structure constants of the Lie algebra, and $\delta(x-y)$ is a delta-function, which in the case of a circle $S^1$ should be understood as a $\mathrm{mod}\; 2\pi$ delta-function. The Poisson structure is naturally continued to arbitrary functions of the fundamental variables $m(x)$.

The Landau-Lifshitz Hamiltonian is defined as follows:
\bea\label{LLham}
H_{LL}=\frac{1}{2} \int\,dx\;\tr(\dd_x m)^2 .
\eea
One can then easily show that Hamilton's equations, which follow from (\ref{poisson}) and (\ref{LLham}), have the following form:
\bea\label{LLeq}
\dd_t m = [m,\dd_x^2 m].
\eea
The important point to realize is that (\ref{LLeq}) is an equation on a function $m$ taking values in a vector space $\mathfrak{g}$ (which is also a Lie algebra of course), which means that it does not satisfy any additional constraints. However, it is easy to see that the time evolution of eq. (\ref{LLeq}) preserves the following local quantities
\bea
l_k(x)=\tr(m(x)^k),\quad k=1, 2, 3...
\eea
Note that only a finite number of $l_k(x)$ are independent quantities.
It follows that the motion actually takes place on proper submanifolds of $\mathfrak{g}$. These submanifolds are parametrized by the values of $l_k(x)$, and on each such manifold the Poisson bracket (\ref{poisson}) induces a symplectic form, so that each such manifold is symplectic. This is in contrast to the original manifold (vector space) $\mathfrak{g}$, which was Poisson but not symplectic (in fact there are plenty of cases when it is odd-dimensional). The symplectic submanifolds mentioned above are also called symplectic leaves of the corresponding Poisson structure.

A particularly interesting case is when $l_k(x)$ are constants, i.e. when they do not depend on $x$. If one chooses particular values for these constants, one obtains the various coadjoint orbits of the group $G$, and the equation (\ref{LLeq}) then naturally reduces to these particular orbits. For example, for the Lie algebra $\mathfrak{g}=\mathfrak{so}_3$ the only nontrivial orbit is the two-sphere, corresponding to $l_2(x)= R^2>0$, and in this case (\ref{LLeq}) turns into the Heisenberg ferromagnet equation (\ref{heisferr}).

\section{Proof that the embedding $\cf_N\hookrightarrow (\CP^{N-1})^{\times N} $\\ is Lagrangian and isometric.}
\label{appflag}

In this Section we first explain what the manifold of complete flags is, and then we show that it is possible to embed the complete flag $\cf_N=U(N)/U(1)^N$ into $(\CP^{N-1})^N$ isometrically and, moreover, as a Lagrangian submanifold.

\subsection{The manifold of complete flags $\cf_N$}

Let $V$ be a vector space. A complete flag (or simply a flag in what follows) is by definition a sequence $V_1, ..., V_m=V$ of vector subspaces of $V$, with the property $V_i \subset V_{i+1}$ and $\textrm{dim} \,V_i =i$. A flag may be given by an (ordered) sequence of $m$ linearly independent vectors $(v_1, ... , v_m)$. Here $v_1$ generates $V_1$, $(v_1, v_2)$ generate $V_2$, etc. The group $GL(m)$ acts transitively on the space of such flags. Indeed, denote the matrix $(v_1, ... , v_m)$ by $F_1$ and $(w_1, ... , w_m)$ by $F_2$. Since the vectors in both sets are linearly independent, the two matrices are nondegenerate. Thus, we can pick $g=  F_1^{-1} F_2 \in GL(m)$, which brings one flag to the other one: $F_2 = F_1 g$. However, some of the matrices $g$ preserve a given flag. After a moment's thought one realizes that these are the upper triangular matrices $g\in B$ (where $B$ stands for the Borel subgroup, which we identify with the upper triangular matrices). Indeed, for upper triangular $g$ the right action $(v_1, ... , v_m)\circ g$ multiplies $v_1$ by a scalar, thus preserving $V_1$, produces a linear combination of $v_1$ and $v_2$ instead of the original $v_2$, thus leaving $V_2$ unaltered, etc. This means that the flag space $\mathcal{F}$ is homeomorphic to $GL(m)/B$.

In what follows we will view the space $U(N)/U(1)^N$ as the space of $N$ ordered lines in $\mathbf{C}^N$, orthogonal with respect to the scalar product $(a,b)\equiv \sum\limits_{i=1}^N a_i^\ast b_i$. One can check that this is equivalent to the definition given above.

\subsection{The embedding is Lagrangian.}

Let \(y\) be a point of the complete flag, that is \(y\equiv\{x_1,...,x_N\}\) is a set of \(N\) complex vectors (in $\CC^N$, i.e. with \(N\) components each) representing the \(N\) orthogonal lines. This means, in particular, that  each of the vectors is defined up to multiplication by a complex number, or in other words $y \in (\CP^{N-1})^{\times N}$. It follows that there's an embedding $\cf_N \hookrightarrow (\CP^{N-1})^{\times N}$. Our statement is as follows: this embedding is Lagrangian. The latter statement comprises two facts:

\begin{itemize}
\item 1) the 2-form induced on \(\cf_N\) by this embedding is identically zero \textbf{and}
\item 2) the dimension of \(\cf_N\) is half the dimension of \((\CP^{N-1})^{\times N}\).
\end{itemize}

Clearly, the second statement is easier to verify. Indeed, since $\cf_N=U(N)/U(1)^N$, its dimension over the real numbers is \(\mathrm{dim}\;\cf_N=N^2-N\), whereas \(\mathrm{dim}_\mathbf{R}\,\CP^{N-1}=2(N-1)\), and it follows that \(\mathrm{dim}\; [(\CP^{N-1})^{\times N} ]=2(N-1)N=2\, \mathrm{dim}\;\cf_N \).

Let us now prove the first statement for the special case of $N=3$. It is straightforwardly generalized to the case of arbitrary $N$. When $N=3$ we have three vectors, which we will call $a, b, c$. The orthonormality conditions for these vectors looks as follows:
\bea\label{orth}
a_i \bar{b}_i=0,\quad a_i \bar{c}_i=0,\quad b_i \bar{c}_i=0.
\eea
The symplectic form on the ambient space $(\CP^2)^{\times 3}$ is
\bea\label{sympl}
\Omega=\omega_{FS}(a)+\omega_{FS}(b)+\omega_{FS}(c),
\eea
meaning that in each term the Fubini-Study form should be taken in the \(a\), \(b\) or \(c\) coordinates. An important thing to realize is that the form \(\Omega\) being zero on the flag $\cf_3$ means that it gives zero when acting on any two vector fields tangent to $\cf_3$: \(\Omega(t_1,t_2)=0\) for $t_{1,2}\in \Gamma(T\cf_3)$. This is what we are going to prove, since choosing an explicit parametrization for the flag manifold is not a very easy enterprise. In practice we will use an overcomplete system of tangent vectors, which may be written as 
\bea\label{r1}
v^n(a,b,c)=w^n(a)+w^n(b)+w^n(c),
\eea
where
\bea\label{r2}
w^n(x)=(\lambda^n)_{ij} \bar{x}^i \frac{\partial}{\partial \bar{x}^j}-(\lambda^n)_{ji} x^i \frac{\partial}{\partial x_j},\quad n=1...9
\eea
and $\lambda^n$ are the Gell-Mann matrices and the identity matrix. Since there are nine tangent vectors to a six-dimensional flag manifold $\cf_3$, not all of them are linearly independent. This does not come as any surprise, since some of these vector fields (or perhaps their linear combinations) act along the orbit of the denominator $U(1)^3$ of the coset $U(3)/U(1)^3$.

It is easy to check that the vector fields (\ref{r1}-\ref{r2}) annihilate the defining conditions (\ref{orth}) of the flag manifold, which means precisely that they are tangent to it. As explained above, we will now evaluate the two-form $\Omega$ on these vectors, i.e. we will calculate $\Omega(v^n,v^m)$ and prove that this is zero.

Using (\ref{sympl}), we obtain
\bea
\Omega(v^n,v^m)=\omega_{FS}(w^n(a),w^m(a))+\omega_{FS}(w^n(b),w^m(b))+\omega_{FS}(w^n(c),w^m(c))
\eea
The Fubini-Study form (\ref{FS}) has two terms. Let us evaluate each of them separately, starting from the second one:
\bearr
&&\frac{da_i \bar{a}_i \wedge d\bar{a}_k a_k}{(\bar{a}_j a_j)^2}(w^n(a),w^m(a))=\\&&=\frac{1}{(\bar{a}_j a_j)^2}\big[da_i(w^n) \bar{a}_i \, d\bar{a}_j(w^m) a_j - da_i(w^m) \bar{a}_i \, d\bar{a}_j(w^n) a_j\big]=\\
&&=\frac{1}{(\bar{a}_j a_j)^2}[-(\bar{a} \lambda^n a)(\bar{a} \lambda^m a)-(m\leftrightarrow n)]=0
\eearr
The first term in the Fubini-Study form produces the following after evaluation:
\bearr
\frac{da_i \wedge d\bar{a}_i}{\bar{a}_j a_j}(w^n(a),w^m(a))=\frac{1}{(\bar{a}_j a_j)}\big[da_k(w^n) d\bar{a}_k(w^m)-da_k(w^m) d\bar{a}_k(w^n)\big]=\\=\frac{1}{(\bar{a}_j a_j)}\big[-\bar{a} \lambda^n\lambda^m a-(m\leftrightarrow n)\big]=\frac{1}{(\bar{a}_j a_j)} \bar{a} [\lambda^m, \lambda^n] a
\eearr
Summing this over \(a\), \(b\) and \(c\), we obtain:
\bea\label{i1}
\Omega(v^n,v^m)=\frac{1}{(\bar{a}_j a_j)} \bar{a} [\lambda^m, \lambda^n] a+\frac{1}{(\bar{b}_j b_j)} \bar{b} [\lambda^m, \lambda^n] b+\frac{1}{(\bar{c}_j c_j)} \bar{c} [\lambda^m, \lambda^n] c
\eea
Since \(a, b, c\) are orthogonal to each other,
\bea\label{i2}
\frac{\bar{a}\otimes a}{(a,a)}+\frac{\bar{b}\otimes b}{(b,b)}+\frac{\bar{c}\otimes c}{(c,c)}=I_3
\eea
is a unit matrix, and it follows from (\ref{i1}-\ref{i2}) that
\bea
\Omega(v^n,v^m)=\tr([\lambda^m, \lambda^n])=0
\eea
The proof is complete.

\subsection{The embedding is isometric.}

Similarly to the case of symplectic structures, which was discussed in Section \ref{aff1}, the metric on the direct product of two manifolds is naturally the sum of metrics on each factor. Our second claim is that the very same embedding \(\cf_N\hookrightarrow (\CP^{N-1})^{\times N}\) is isometric.

We start the proof by considering the metric entering the action (\ref{flagaction}):
\bear\label{fm1}
&ds^2= |u_1 \circ d \bar{u}_2|^2+|u_1 \circ d \bar{u}_3|^2+|u_2 \circ d \bar{u}_3|^2,\quad \textrm{where}&\\ \label{fm2}
&u_i \circ \bar{u}_j = \delta_{ij}&
\eear
It can clearly be rewritten as follows:
\bea
ds^2={1\over 2}\left( du_1 \left[\bar{u}_2 \otimes u_2 + \bar{u}_3 \otimes u_3\right] d\bar{u}_1 +
du_2 \left[\bar{u}_1 \otimes u_1 + \bar{u}_3 \otimes u_3\right] d\bar{u}_2+
du_3 \left[\bar{u}_1 \otimes u_1 + \bar{u}_2 \otimes u_2\right] d\bar{u}_3
\right)
\eea
Now we apply (\ref{i2}) to each of the terms in the square brackets to obtain:
\bea
ds^2={1\over 2}\left( du_1 \;(\mathrm{I}-\bar{u}_1 \otimes u_1)\; d\bar{u}_1 +
du_2 \;(\mathrm{I}-\bar{u}_2 \otimes u_2)\; d\bar{u}_2+
du_3 \;(\mathrm{I}-\bar{u}_3 \otimes u_3)\; d\bar{u}_3 \right),
\eea
where $\mathrm{I}$ is the $3\times 3$ identity matrix. It is clear that each term of the form $du \;(\mathrm{I}-\bar{u} \otimes u)\; d\bar{u}$ is the Fubini-Study metric written in the ``semi-gauge'' $|u|=1$. It follows immediately that the metric (\ref{fm1}) is the sum of three Fubini-Study metrics restricted by the conditions (\ref{fm2}):
\bea
g_{\cf_3}=\;\;\textrm{Restriction of}\;\;\;\;\;{1\over 2} (g^{(1)}_{FS}+g^{(2)}_{FS}+g^{(3)}_{FS}).
\eea
The proof is complete.

\bibliography{refs}
\bibliographystyle{alpha}

\end{document}